\documentclass[usenatbib]{mn2e}
\usepackage{times}
\usepackage{graphicx}
\usepackage{amsmath}
\usepackage{mathabx}
\usepackage{subfigure}
\usepackage{natbib}
\usepackage[colorlinks=true, linkcolor=blue, citecolor=blue, urlcolor=blue]{hyperref}
\usepackage{txfonts}

\newcommand{\aj}{AJ}
\newcommand{\mnras}{MNRAS}
\newcommand{\apj}{ApJ}
\newcommand{\apjs}{ApJS}
\newcommand{\apjl}{ApJL}
\newcommand{\aap}{A\&A}
\newcommand{\aaps}{A\&AS}
\newcommand{\apss}{ApSS}
\newcommand{\nat}{Nature}
\newcommand{\pasp}{PASP}

\newcommand{\araa}{ARA\&A}
\newcommand{\iaucirc}{IAU\ Circ.}
\newcommand{\na}{NewA}

\defcitealias{1999A&A...344..868X}{X99}	
\defcitealias{2007A&A...471..765B}{B07}

\begin{document} 

\title[Dust in edge-on spiral galaxies]{The distribution of interstellar dust in CALIFA edge-on galaxies via oligochromatic radiative transfer fitting}

\author[G. De Geyter, M. Baes, P. Camps]
{Gert De Geyter,$^1$ 
Maarten Baes,$^1$ 
Peter Camps,$^1$
Jacopo Fritz,$^1$ 
Ilse De Looze,$^1$ 
\newauthor
Thomas M. Hughes,$^1$ 
S\'ebastien Viaene, $^1$
Gianfranco Gentile$^{1,2}$\\
$^1$Sterrenkundig Observatorium, Universiteit Gent, Krijgslaan 281-S9, B-9000 Gent, Belgium \\
$^2$Department of Physics and Astrophysics, Vrije Universiteit Brussel, Pleinlaan 2, 1050 Brussels, Belgium}

\date{\today}
\maketitle
\begin{abstract}
We investigate the amount and spatial distribution of interstellar dust in edge-on spiral galaxies, using detailed radiative transfer modeling of a homogeneous sample of 12 galaxies selected from the CALIFA survey. Our automated fitting routine, FitSKIRT, was first validated against artificial data. This is done by simultaneously reproducing the SDSS $g$-, $r$-, $i$- and $z$-band observations of a toy model in order to combine the information present in the different bands. We show that this combined, oligochromatic fitting, has clear advantages over standard monochromatic fitting especially regarding constraints on the dust properties. 

We model all galaxies in our sample using a three-component model, consisting of a double exponential disc to describe the stellar and dust discs and using a S\'ersic profile to describe the central bulge. The full model contains 19 free parameters, and we are able to constrain all these parameters to a satisfactory level of accuracy without human intervention or strong boundary conditions. Apart from two galaxies, the entire sample can be accurately reproduced by our model. We find that the dust disc is about 75\% more extended but only half as high as the stellar disc. The average face-on optical depth in the V-band is $0.76$ and the spread of $0.60$ within our sample is quite substantial, which indicates that some spiral galaxies are relatively opaque even when seen face-on. 
\end{abstract} 

\section{Introduction}

A part of the interstellar medium (ISM) is made up from dust grains, created mainly in the atmospheres of AGB stars \citep{2006A&A...447..553F, 2011ApJ...727...63D} and during supernova explosions \citep{1998ApJ...501..643D, 2010ASPC..425..237C, 2014ApJ...782L...2I}. Apart from their role in a number of physical and chemical processes, like catalyzing the production of molecular hydrogen and consequently regulating star formation \citep{1971ApJ...163..155H}, dust grains are important as they absorb at least one third of the UV/optical light in galaxies \citep{2002MNRAS.335L..41P}. 

By using far-infrared (FIR) and submm observations, it has become possible to determine the total dust mass in galaxies with a reasonable accuracy.  Especially since the launch of the {\it Herschel} Space Observatory \citep{2010A&A...518L...1P}, the dust mass can be directly estimated by fitting a modified blackbody or more advanced models to the observed spectral energy distribution \citep[e.g.,][]{2011MNRAS.417.1510D, 2012ApJ...745...95D, 2013MNRAS.436.2435S, 2014arXiv1402.3597C}. However, apart from a few nearby targets, these observations lack the spatial resolution to resolve the individual regions and getting a more detailed picture of the dust distribution.

An alternative approach consists of determining the dust masses and distribution from the extinction effects caused by dust on UV and optical starlight. In this case, the necessary resolution is easily attained and the observations are cheap and easy to obtain. The drawback is that it is not straightforward to determine the total mass and the distribution of the dust from the observed level of extinction \citep{1989MNRAS.239..939D, 1992ApJ...393..611W, 1994ApJ...432..114B, 1995MNRAS.277.1279D, 2001MNRAS.326..733B, 2005MNRAS.359..171I} . To adequately determine these properties from extinction all the necessary physics like the absorption efficiency, scattering rate, etc. for different dust compositions and distributions have to be accounted for. In other words, one has to rely on dust radiative transfer modeling. 

Thanks to an increase in computational power and the development of efficient algorithms, dust radiative transfer codes have become increasingly more realistic and powerful. They can now be applied to arbitrary geometries, and include physical processes as absorption, multiple anisotropic scattering, polarization, thermal and non-thermal emission \citep[e.g.][]{2001ApJ...551..269G, 2003MNRAS.343.1081B, 2011ApJS..196...22B, 2006MNRAS.372....2J, 2008A&A...490..461B, 2011A&A...536A..79R}. An overview of the most important developments, advantages and disadvantages of different approaches regarding 3D dust radiative transfer codes can be found in \citet{2013ARA&A..51...63S}.

As the creation of these radiative transfer models of spiral galaxies can be an arduous task, most of these studies focus on a very limited number of targets or even a single galaxy  \citep[e.g.][]{1987ApJ...317..637K, 2000A&A...362..138P, 2011A&A...527A.109P, 2000A&A...359...65B, 2001ApJ...548..150M, 2001A&A...372..775M, 2010A&A...518L..39B, 2011ApJ...741....6M, 2012MNRAS.419..895D, 2012MNRAS.427.2797D, 2012ApJ...746...70S}. Most of these studies have targeted spiral galaxies with an inclination close to 90 degrees, as these offer some unique advantages. In particular, because of the edge-on projection, the separate components of the galaxy, i.e.\ the stellar disc, dust disc and central bulge, are clearly distinguishable. Unfortunately, the varying observational input used for the modeling, and the different assumptions, geometries and physical ingredients in the radiative transfer models themselves make it almost impossible to compare these studies and draw general implications on the amount and distribution of dust in spiral galaxies. What is needed is a radiative transfer study of a sizable sample of galaxies, all modeled in a homogeneous way based on similar data. 

There have been only two such efforts in the literature: \citet[][hereafter \citetalias{1999A&A...344..868X}]{1997A&A...325..135X, 1998A&A...331..894X, 1999A&A...344..868X} and \citet[][hereafter \citetalias{2007A&A...471..765B}]{2007A&A...471..765B} both modeled a set of seven edge-on spiral galaxies with prominent dust lanes. The main conclusions of these works are that the dust disc tends to be larger radially than the stellar disc while being thinner in the vertical direction. However, there is still some uncertainty on the ratio of the dust and stellar scale length as different methods to determine the size of the dust disc results in values from being from only 10\% larger \citep{2009ApJ...701.1965M} up to 100\% or even more \citep{2005AJ....129.1396H, 2010A&A...509A..91T, 2011ApJ...741....6M}. In the cases where the FIR/submm observations were able to resolve the dust disc, similar values for the dust scale height and length were found as the ones determined from fitting radiative transfer models to optical/NIR observations \citep{2013A&A...556A..54V, 2014arXiv1402.5967H}.

One of the most important quantities concerning the dust distribution in spiral galaxies is the face-on optical depth  $\tau_{\text{V}}^{\text{f}}$. A large number of papers looked into this parameter, resulting in often contradicting conclusions \citep{1990Natur.346..153V, 1994MNRAS.266..614V, 1991Natur.353..515B, 1992ApJ...400L..21B, 1993PASP..105..993B, 1995ApL&C..31..143B, 1998A&A...334..772B, 2003AJ....126..158M}. The importance of this value is not to be underestimated: galaxies with $\tau_{\text{V}}^{\text{f}}<1$ would be almost completely transparent when seen face-on, whereas higher values for $\tau_{\text{V}}^{\text{f}}$ would indicate that a significant fraction of the stars in spiral galaxies could be invisible. The latter case has far-reaching consequences for the use of spiral galaxies in a cosmological context, as it would, for example, invalidate most determinations of mass-to-light ratios.  

Another important aspect to consider is the ratio between the stellar and dust scale height. According to \citet{2004ApJ...608..189D} there should be a clear transition at the critical rotation speed of 120 km\,s$^{-1}$ where the more massive galaxies have a well-defined, thin dust disc, while the less massive galaxies have a more diffuse and bloated dust disc. A similar feature was also observed by \citet{2011ApJ...741....6M} whose radiative transfer models contain a dust disc which has a scale height similar to or larger than the stellar disc and therefore does not have an obvious dust lane, contrary to the relatively much thinner dust discs in galaxies with a higher surface brightness. They suggest that the origin might be a combination of the two possible explanations: either their galaxies have a larger stability against vertical collapse, as expected using the results described in \citet{2004ApJ...608..189D}, or the extraordinary thin nature of the stellar discs in their sample. 

In this paper we study a set of edge-on spiral galaxies by using detailed radiative transfer models in order to determine their stellar and dust morphology and, subsequently, their face-on optical depth. The selection of our sample, the largest sample so far upon which detailed radiative transfer modeling has been applied, is discussed in Sect.~{\ref{SampleSelection.sec}}. In Sect.~{\ref{RT.sec}} we present our modeling technique, which should be objective and preferably automated. We use an updated version of FitSKIRT \citep{2013A&A...550A..74D}, an automated optimizing code around the radiative transfer code SKIRT \citep{2011ApJS..196...22B} to construct models which accurately reproduce the observations. The most important new feature of the code is the oligochromatic fitting\footnote{oligo- or olig-, as in oligopoly, oligarchy or oligosaccharide, derived from the Greek {\em{ol{\'\i}gos}}, meaning "few" or "a little". So oligochromatic fitting is the modeling of a small number of images simultaneously.}, which is discussed and thoroughly tested on an artificial data. The results of our fits and an individual discussion of each galaxy in our sample is presented in Sect.~{\ref{Results.sec}}. We make some remarks about the quality of the models in general, compare and validate our results against results available from other studies, and discuss the results of both the stellar and dust properties of our sample in Sect.~{\ref{Discussion.sec}}. Our conclusions are presented in Sec.~{\ref{Conclusions.sec}}.

\section{Sample selection}
\label{SampleSelection.sec}

Our starting point was the Calar Alto Legacy Integral Field Area Survey (CALIFA) sample \citep{2012A&A...538A...8S}. This is a statistically well-defined sample of 600 nearby galaxies, selected from the Sloan Digital Sky Survey (SDSS) DR7 photometric catalog \citep{2009ApJS..182..543A} by using two additional conditions: the redshift range was limited to $0.005 < z < 0.03$ and only galaxies with an isophotal $r$-band diameter between 45 and 80 arcsec were considered. The CALIFA survey is one of the largest integral field spectroscopic surveys: by completion of the survey, all galaxies will have been observed using the PMAS/PPAK integral field spectrograph \citep{2005PASP..117..620R, 2006PASP..118..129K} in the wavelength range between 3700 and 7000~{\AA}. This will allow the construction of detailed two-dimensional maps of the stellar kinematics, the ages and metallicities of the stellar populations, and the distribution, excitation mechanisms, chemical abundances and kinematics of the ionized gas. The first CALIFA public data release took place in November 2012 \citep{2013A&A...549A..87H} and several scientific results based on these data have been published \citep[e.g.,][]{2012A&A...540A..11K, 2013A&A...554A..58S, 2013arXiv1311.7052S, 2013A&A...559A.114M, 2014A&A...562A..47G}. The size of the sample and the wealth of information that is (or will be) available for all galaxies in a homogeneous way make it an ideal starting sample from which we select ours. 

\begin{figure*}
\centering
\includegraphics[width=\textwidth]{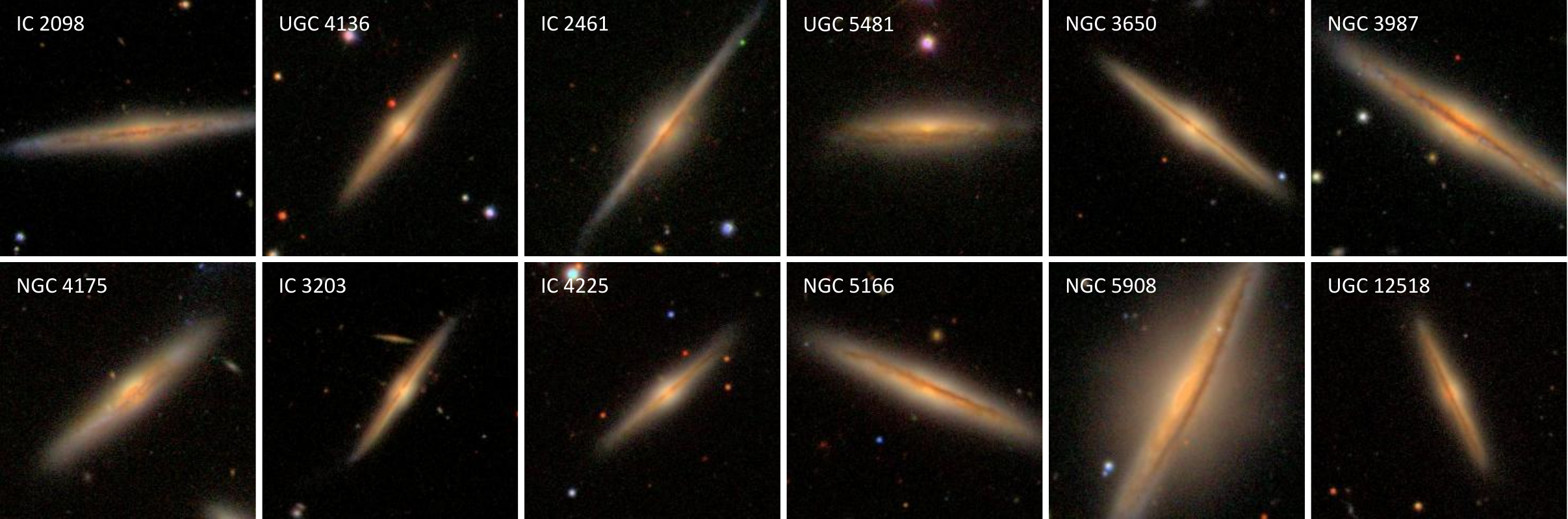} 
\caption{SDSS thumbnail images for the 12 galaxies in our sample with a FoV of $90"$ by $90"$.}
\label{Sample.fig}
\end{figure*}

From the CALIFA sample, edge-on spiral galaxies with a clear dust lane morphology were selected by visual inspection of the SDSS images of the galaxies. Interacting or visibly strongly asymmetrical galaxies were rejected from the sample. In order to do a reliable radiative transfer fitting (the goal of this paper), it is crucial that the galaxies have sufficient resolution. In particular, if we want to derive information on the distribution of the dust, the dust lane must be properly resolved in the vertical direction. Based on previous experience with the fitting code, FitSKIRT \citep{2013A&A...550A..74D}, we exclude galaxies with a major axis smaller than 1 arcmin or a minor axis smaller than 8 arcsec. The final sample, listed in Table \ref{Sample.tab}, consists of 12 edge-on spiral galaxies with distances ranging from 42 to 119 Mpc. Figure~{\ref{Sample.fig}} shows SDSS thumbnail images of the sample galaxies.

Apart from their statistical properties, an added value of these galaxies over other SDSS galaxies is the fact that their integral field spectroscopy might be used in follow-up studies. The H$\alpha$/H$\beta$ line ratio could be used to have an independent measure of the attenuation in our sample \citep{2004A&A...419..821T, 2013A&A...550A.114B}. Additionally the CALIFA data might be used to determine the gas-phase metallicity gradients across star-forming disks \citep{2013A&A...554A..58S}. Combining maps of the metallicity distribution with our ancillary data to simultaneously trace the gas, stellar and dust content, enables us to study the relationships between the key phases of the baryons in the ISM. The relationship between stellar mass, oxygen abundance and gas content found in large scale, integrated observations \citep{2013A&A...550A.115H, 2013MNRAS.433.1425B} could therefore be tested on local scales. The predicted relationship between the gas-to-dust ratio and the gas-phase metallicity may also be used to constrain theoretical models of chemical evolution \citep{2007ApJ...663..866D}. Furthermore, we can directly test for the accretion of cold gas by searching for pristine, unenriched gas in the exterior of the disks \citep{2009ApJ...695..580B}. We therefore have many opportunities for follow-up studies.

For all sample galaxies, homogeneous $u$, $g$, $r$, $i$ and $z$-band images are available from the SDSS, so, in principle, we could use these five bands for our multi-wavelength radiative transfer fits. However, we decided not to include the $u$-band images, as the image quality was insufficient for most galaxies. Additional near-infrared $J$, $H$ and $K_s$ imaging is available for all galaxies from the Two Micron All Sky Survey \citep[2MASS,][] {2006AJ....131.1163S}, but also here the signal-to-noise ratio was inadequate for most galaxies in the sample. An alternative could be deep NIR imaging from the UKIRT Infrared Deep Sky Survey \citep[UKIDSS,][]{2007MNRAS.379.1599L}, but this is available only for a few galaxies in the sample. As we wanted to keep the data set homogeneous, we did not include these data. 

This complete and uniform sample of $g$, $r$, $i$ and $z$-band data is an ideal set to investigate the dust properties in extinction in a systematic way: the $g$ band is clearly attenuated by the dust while the $z$ band observations contain more accurate information about the stellar properties.

\begin{table*}
\centering
\begin{tabular}{lccccccc}
\hline\hline\\[-1ex]
Galaxy & RA & Dec & Distance & Major axis & Minor axis & Type & $r$-band\\
 & (J2000) & (J2000) & (Mpc) & (arcmin) & (arcmin) & & (mag)\\[2ex]
 \hline \\[-1ex]
IC\,2098 & 04:50:44 & $-$05:25:07 & 47.8 & 2.3 & 0.3 & Sc & 13.8\\
UGC\,4136 & 07:59:54 & +47:24:47 & 90.8 & 1.7 & 0.4 & Sa & 13.5\\
IC\,2461 & 09:19:58 & +37:11:29 & 54.2 & 2.4 & 0.4 & Sbc & 13.3\\
UGC\,5481 & 10:09:51 & +30:19:15 & 90.2 & 1.5 & 0.5 & Sa & 13.1\\
NGC\,3650 & 11:22:35 & +20:42:14 & 59.7 & 1.7 & 0.3 & SAb & 13.4 \\
NGC\,3987 & 11:49:30 & $-$01:05:11 & 61.4 & 2.2 & 0.4 & Sb & 13.8 \\
NGC\,4175 & 12:12:31 & +29:10:06 & 42.2 & 1.8 & 0.4 & Sbc & 13.2 \\
IC\,3203 & 12:21:46 & +25:53:05 & 119.2 & 1.5 & 0.2 & Sb & 14.2 \\
IC\,4225 & 13:20:01 & +31:58:53 & 73.9 & 1.3 & 0.3 & S0/a & 13.9 \\
NGC\,5166 & 13:28:15 & +32:01:57 & 62.6 & 2.3 & 0.4 & Sb & 12.9\\
NGC\,5908 & 15:16:43 & +55:24:33 & 53.4 & 3.2 & 1.2 & SAb & 12.0\\
UGC\,12518 & 23:20:13 & +07:55:55 & 51.6 & 1.3 & 0.2 & Sb & 14.0 \\[2ex]
\hline\hline
\end{tabular}
\caption{The final sample of 12 edge-on galaxies selected out of the CALIFA sample.}
\label{Sample.tab}
\end{table*}

\section{Radiative transfer modeling}
\label{RT.sec}

\subsection{The model}
\label{Model.sec}

The model we use to fit to the image consists of a double-exponential stellar disc, a flattened S\'ersic bulge and double-exponential dust disc -- this is the {\it de facto} standard used for radiative transfer modeling of edge-on spiral galaxies \citep{1999A&A...344..868X, 2000A&A...362..138P, 2011A&A...527A.109P, 2007A&A...471..765B, 2010A&A...518L..39B, 2012MNRAS.427.2797D, 2013A&A...550A..74D}. 

The stellar disc is described by a scale length $h_{R,*}$, a scale height $h_{z,*}$, and a luminosity at each of the $g$, $r$, $i$ and $z$ bands. The bulge is parameterized by the effective radius $R_{\text{eff}}$, the S\'ersic index $n$, the intrinsic flattening $q$, and again, a luminosity at each of the four bands considered. Note that we use the same geometrical parameters at all wavelengths and that only the luminosities of the bulge and disc are determined individually at each wavelength. While there is evidence that e.g.\ disc scale lengths are wavelength-dependent \citep{1994A&AS..108..621P, 1996ASSL..209..523C, 1999A&A...344..868X}, we have chosen this approach to limit the number of free parameters in our models. A geometrical parameter that is expected to change as a function of wavelength is the bulge-to-disc ratio: typically, it increases steadily from blue wavelengths, where young stars in the disc dominate, into the near-infrared, where old stellar populations in the bulge contribute the most. The fact that the individual luminosities of disc and bulge at each band are free parameters in our model, allows us to simulate this wavelength dependent behavior. 

The geometry of the dust disc is also characterized by a scale length $h_{R,{\text{d}}}$ and scale height $h_{z,{\text{d}}}$. Apart from the geometrical distribution of the dust, the optical properties (absorption efficiency, scattering efficiency and scattering phase function) and the total amount of dust need to be set. We have used the standard BARE-GR-S model of \citet{2004ApJS..152..211Z}, which consists of a combination of PAH, graphitic and silicate dust grains. The relative distributions of each component have been weighted to best match the extinction, abundances, and emission associated with the Milky Way dust properties ($R_V = 3.1$). By using a physical dust model, we have fixed the wavelength dependence of the optical properties of the dust, and the only free parameter left to determine is the total amount of dust. This can be characterized in different ways, e.g.\ as the total dust mass $M_{\text{d}}$ or the face-on or edge-on optical depth at any wavelength (for a given geometry and fixed dust model, these parameters are equivalent).

Although stellar discs are often truncated at 3 to 4 scale lengths \citep{2000A&A...357L...1P, 2007MNRAS.378..594P, 2001MNRAS.324.1074D, 2002MNRAS.334..646K}, we have not included an explicit truncation profile for either the dust or stellar disc. For the dust disc, we use a grid to partition the dust distribution that is four times the upper boundary of the dust scale length. This results in an implicit truncation of the dust distribution, but as only a very small fraction of the dust mass falls outside this radius, our discs are essentially not truncated. Note that \citetalias{1999A&A...344..868X} and \citetalias{2007A&A...471..765B} did use a truncation of the dust disc at 3\,--\,4 scale lengths, which could potentially complicate a direct comparison.

The last three parameters to be determined are the inclination $i$ of the system and the x and y coordinate of the centre of the galaxy on the plane of the sky. In total, our model has 19 free parameters (5 stellar geometry parameters, 8 luminosities, 2 dust geometry parameters, the total dust mass and 3 projection parameters). Our problem comes down to finding the combination of these 19 parameters that results in a set of simulated $g$, $r$, $i$ and $z$-band images that reproduce the observed images best. This is extremely time consuming as each simulation requires a full radiative transfer calculation including absorption and multiple anisotropic scattering by dust. In addition, both the observed images and the simulated images contain Poisson noise (which is unavoidable when using a Monte Carlo code for radiative transfer calculations). Moreover, the 19-dimensional parameter space is typically not well-behaved, with different local extrema. Special techniques are therefore required.

\subsection{FitSKIRT}
\label{FitSKIRT.sec}

We use the FitSKIRT radiative transfer modeling code \citep{2013A&A...550A..74D}. The core of FitSKIRT is SKIRT \citep{2011ApJS..196...22B}, a Monte Carlo radiative transfer code designed to solve the 3D dust radiative transfer problem including the physical processes of absorption, multiple anisotropic scattering and thermal emission by dust \citep[for a general overview of radiative transfer, see][]{2013ARA&A..51...63S}. Advanced ingredients of the SKIRT code include the use of hierarchical and/or unstructured grids for the dust medium \citep{2013A&A...554A..10S, 2014A&A...561A..77S, 2013A&A...560A..35C}, a range of optimization techniques to speed up the Monte Carlo process \citep{2008MNRAS.391..617B, 2011ApJS..196...22B}, and the possibility to calculate the effect of dust on the observed kinematics \citep{2001ApJ...563L..19B, 2002MNRAS.335..441B, 2003MNRAS.343.1081B}. It is mainly applied to simulate dusty galaxies \citep[e.g.,][]{2010A&A...518L..45G, 2012MNRAS.427.2797D, 2012MNRAS.419..895D}, but has also been used to model active galactic nuclei \citep{2012MNRAS.420.2756S} and dusty discs around evolved stars \citep{2007BaltA..16..101V, Deschamps14}.

FitSKIRT couples SKIRT to the genetic algorithms-based optimization library GAlib \citep{Wall96}. Genetic algorithms are problem solving systems based on evolutionary principles. They have been applied successfully to a large range of global optimization problems, and they are becoming increasingly popular as a tool for astrophysical applications \citep{1995ApJS..101..309C, 2012arXiv1202.1643R}. Genetic algorithms easily handle noisy objective functions, since they work on a set of solutions rather than iteratively progressing from one point to another \citep{2000ApJ...545..974M, 2003AJ....125.1958L}. This makes them particularly suited for our case. A full description of FitSKIRT can be found in \citet{2013A&A...550A..74D}.

For the present work, we use a novel version of FitSKIRT with significantly extended and improved capabilities. The code has been completely rebuilt and restructured. The new version of FitSKIRT, like the previous version, is completely written in C++, but has now been embedded in the Qt framework \citep{Summerfield2010}. This allows for more flexibility in both the implementation and the usability of the program. The code can read an arbitrary number of images and fit a radiative transfer model to them based on an arbitrary combination of stellar and dust components. Moreover, the genetic algorithm based optimization process has been modified so the radiative transfer simulations are now executed in parallel. This parallelization is handled using Qt multi-threading protocols on shared memory computers and by a message passing interface (MPI) on distributed memory clusters. Consequently, the speed of the code has significantly increased making it an ideal tool to use on larger samples.

\subsection{Oligochromatic fitting}

As blue bands show clear signs of extinction by dust and which conceals information about the intrinsic stellar distribution, the opposite is true for longer wavelengths where a more accurate map of the stellar distribution is shown. However this comes at the cost of a detailed information on the dust properties. Consequently an ideal fitting routine should incorporate this information by simultaneously making use of the two wavelength regimes. 
Therefore, the most considerable addition to FitSKIRT is the possibility of oligochromatic fitting. Although fitting to a set of images at different wavelengths is computationally more demanding, it also holds some clear advantages over monochromatic fitting procedures. Apart from the ability to handle some of the degeneracies monochromatic modeling has to cope with, oligochromatic fitting allows for more stable convergence towards the optimal solution, especially for low S/N observations \citep{2013MNRAS.430..330H, 2013MNRAS.435..623V}.  

In the oligochromatic approach, the parameter values of the model are used to generate a set of images at different wavelengths. In essence, the oligochromatic simulation is therefore nothing more than a set of monochromatic images, all having the same input values. The most straightforward way would be to search the entire 19-dimensional parameter space for the best solution. 

We can, however, use the fact that the radiative transfer problem is linear with respect to the total luminosity. For the monochromatic FitSKIRT fitting as presented by \citet{2013A&A...550A..74D}, we determined the total luminosity of the system as a separate optimization outside the genetic algorithm. Here, we go one step further, and determine the luminosities of the bulge and disc at each individual band outside the genetic algorithm. This is achieved as follows: for a particular dust geometry, we initially set the luminosity of each stellar component equal to a dummy value. We create a set of images for the bulge and disc components separately. The next step consists of convolving each of these images with the PSF of the observed image at the corresponding wavelength. Determining the best fit for the luminosity of the bulge and the disc is then done by summing the two components and looking for the two luminosity values that result in the lowest $\chi^2$ at each wavelength. At every wavelength, this corresponds to a simple 2D minimization, which is done using a straightforward downhill simplex method. As the code fits each wavelength individually, no prior assumptions regarding the stellar population in the disc or bulge have to be made. Once the luminosities are determined, the total $\chi^2$-value of the model is determined as the sum of all wavelength dependent $\chi^2$-values.

An important issue in the oligochromatic fitting is the assignment of relative weights of the different frames to the global $\chi^2$ value. Different noise values, pixel scales or simply a different number of pixels in the frame could cause a strong bias in the global $\chi^2$ value. For example, if one of the observed frames has more pixels to fit or higher counts compared to the other frames, it could easily dominate the resulting $\chi^2$-value. As a result, the oligochromatic fit would essentially become a computationally expensive monochromatic fit. To avoid this situation, the creation of the set of reference images should be done as uniformly as possible. The first step is to rebin the frames, assure the field of view and the pixel scale are the same for all wavelengths. This not only assures the total number of pixels where the $\chi^2$-value is determined to be the same but also assures that the same physical region is being studied. 

In some cases, regions containing foreground stars, hot or dead pixels, etc. might need to be masked. As a final step to minimize the biasing of certain frames, we normalize the total intensity in all frames so the total $\chi^2$ weight should be equal for all frames. 

\subsection{Validation of the oligochromatic fitting}
\label{Validation.sec}

\begin{figure*}
\centering
\includegraphics[width=0.62\textwidth]{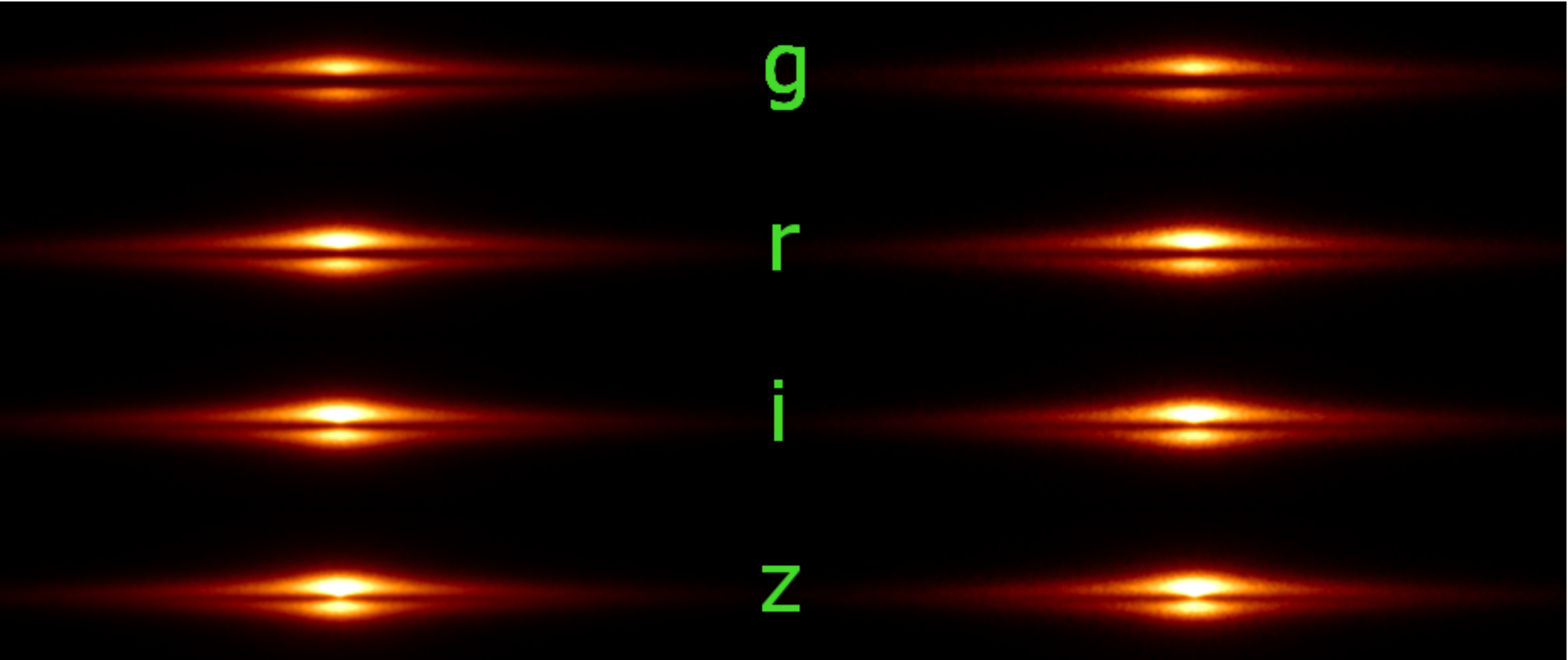} 
\includegraphics[width=0.35\textwidth]{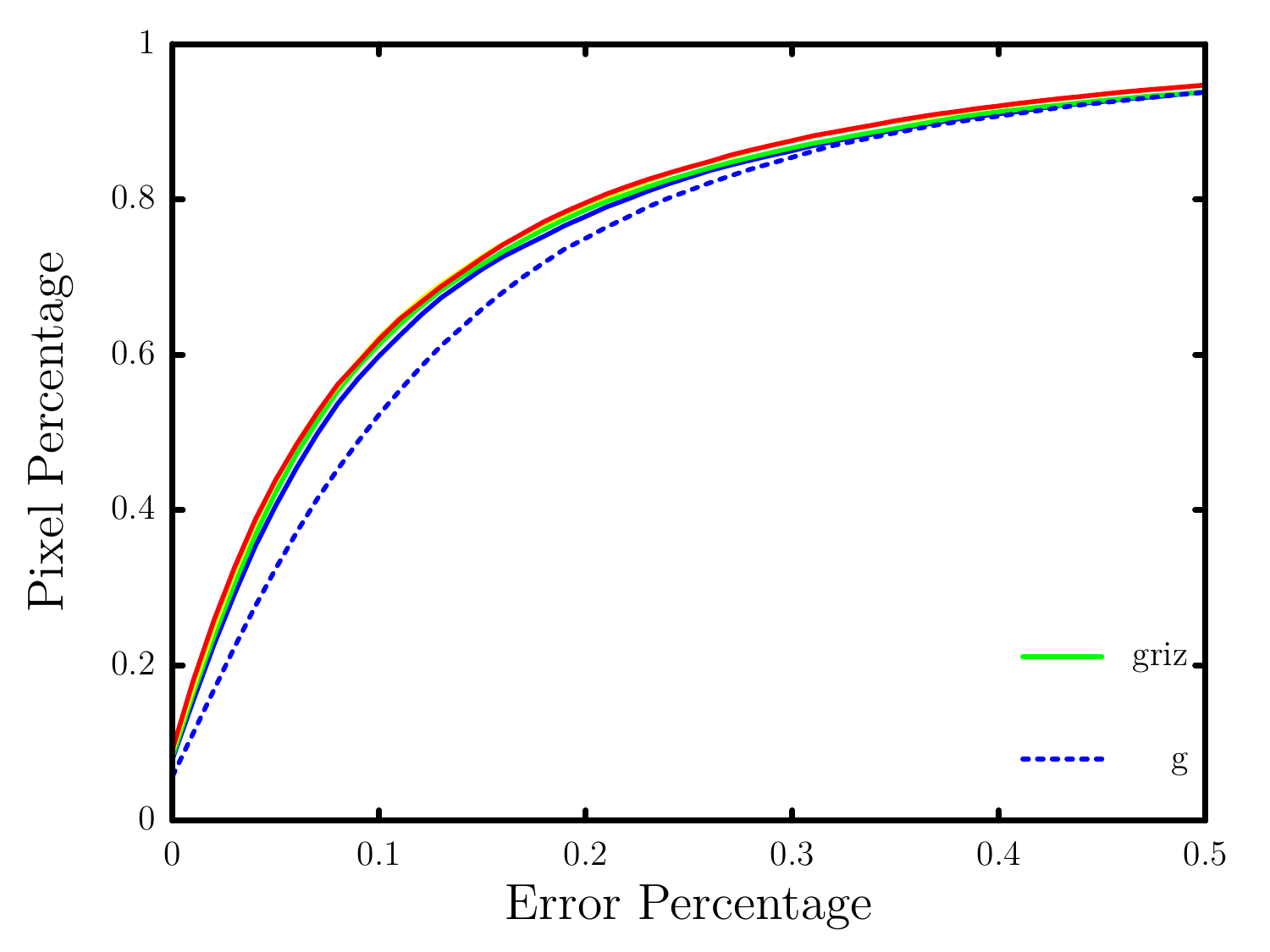}
\caption{{\it Left-hand} panel: Reference frame on the left, best fitting model of an oligochromatic $griz$-fitting on the right. {\it Right-hand} panel: The cumulative pixel percentage distribution within a certain error percentage deviation of the reference frame. The oligochromatic model is shown as solid lines, each one representing a different band, while the $g$-band monochromatic model is shown as a dashed line. The other monochromatic fits are similar to the $g$-band fit.}
\label{Validation.fig}
\end{figure*}

\begin{table}
\centering
\begin{tabular}{ccrr@{$\,\pm\,$}lr@{$\,\pm\,$}l}
\hline\hline\\[-0.5ex]
Parameter & unit & reference & griz & RMS & g & RMS  \\[2ex]
\hline \\[-0.5ex]
$h_{R,*}$ & kpc & 4.4 & 4.36 & 0.11 & 4.46 & 0.17  \\
$h_{z,*}$ & kpc & 0.5 & 0.51 & 0.02 & 0.50 & 0.02 \\
$L_{g}^{\text{tot}}$ & $10^9~L_{\odot}$ & 1.33 & 1.30 & 0.12 & 1.31 & 0.12\\
$R_{\text{eff}}$ & kpc & 2 & 2.2 & 0.7 & 2.1 & 0.8 \\
$n$ & --- & 2.5 & 2.3 & 0.7 & 2.1 & 1.2 \\
$q$ & --- & 0.5 & 0.51 & 0.03 & 0.48 & 0.04 \\
$B/T_{g}$ & --- & 0.21 & 0.19 & 0.04 & 0.22 & 0.05 \\
$h_{R,\text{d}}$ & kpc & 6.6 & 6.5 & 0.6 & 6.4 & 1.1 \\
$h_{z,\text{d}}$ & kpc & 0.25 & 0.25 & 0.01 & 0.24 & 0.03 \\
$M_{\text{d}}$ & $10^7~M_\odot$ & 4 & 4.2 & 0.4 & 3.9 & 0.6 \\
$i$ & deg & 89 & 89.1 & 0.1 & 89.0 & 0.3 \\[2ex]
 \hline\hline
 \end{tabular}
 \caption{Reference values, Best values and RMS values for the mock galaxy determined FitSKIRT results. The RMS is calculated in the same way as done in \citet{2013A&A...550A..74D} }
\label{Validation.tab}
\end{table}

The monochromatic version of FitSKIRT was thoroughly validated, as described in \citet{2013A&A...550A..74D}. As the oligochromatic approach substantially differs from the monochromatic, we revisit the test case used for the latter. An artificial galaxy model consisting of an exponential stellar disc, a flattened S\'ersic bulge and an exponential dust disc was set up using typical values for the various parameters. The model is similar to the test case used in \cite{2013A&A...550A..74D}, with the main exception that we now generate images at the $g$, $r$, $i$ and $z$ bands using the SKIRT code. As a sanity check, we can use this artificial galaxy to look at both the reproduction of the images as well as recovering the initial input values. To set the values of the luminosities of the two components, we assumed that the bulge consists of an old stellar population with an age of 12 Gyr, while the disc is characterized by a young stellar component of 1 Gyr. They were normalized such that the intrinsic bulge-to-disc ratio is 0.33 and the disc luminosity is $10^9~L_{\odot}$ in the V band, corresponding to values of 0.27, 0.37, 0.42 and 0.50  for the bulge-to-disc ratios and $1.33$- ,$1.23$- , $0.97$- and $0.80 \times 10^9~L_{\odot}$ for the disc luminosity in the $g$, $r$, $i$ and $z$ band, respectively. The left column of the left-hand panel in Figure~{\ref{Validation.fig}} shows the resulting $g$, $r$. $i$ and $z$ band images at an inclination of 88 degrees. Both the increasing bulge-to-disc ratio and the decreasing level of extinction with increasing wavelength are clearly visible.

The right column of the left-hand panel in Figure~{\ref{Validation.fig}} shows the result of an oligochromatic FitSKIRT fit to this set of four images. The only clear difference between the corresponding left and central column images is a lower signal-to-noise in the model images. This is the result of the significantly lower number of photon packages used within the FitSKIRT fitting routine. Indeed, as each FitSKIRT fit typically runs several tens of thousands of individual SKIRT radiative transfer calculations, it is clear that one has to find a balance between run time and noise suppression \citep{2013A&A...550A..74D}. The most important conclusion of this test calculation is that all of the free parameters in the model are recovered within the uncertainties. The input parameters and values recovered by both the $g$-band monochromatic fit and the $g$-, $r$-, $i$- and $z$-band oligochromatic fit are listed in Table \ref{Validation.tab}. The errors are determined by using the same approach as discussed in \cite{2013A&A...550A..74D}. f

In order to investigate the strength of the new fitting routine, we also compared the result of our oligochromatic FitSKIRT fit to the results of individual monochromatic FitSKIRT fit. The right panel in Figure~{\ref{Validation.fig}} shows the distribution of cumulative pixel percentage in the residual frames for both the oligochromatic $g$, $r$, $i$ and $z$ images and the monochromatic $g$-band image (the other monochromatic fits have residuals which are comparable to the $g$-band fit). To avoid the low signal-to-noise pixels and focus on the relevant ones we only include the 25,000 most central pixels of the $501\times101$ pixels frames. We note that both fitting methods produce models where about 80\% of the pixels have a deviation within 25\% of the reference image. It is also clear that the oligochromatic fit does no bias any band over the other which is a nice validation of our precautions. It has to be noted, however, that in this case all the images have the same signal-to-noise, which is usually not the case of realistic data sets.

The oligochromatic models have smaller error bars on most of the fitted model parameters than the individual monochromatic models. This is particularly true for the parameters of the dust disc, where the oligochromatic fitting clearly has an added value. Even for the $g$ band image, where the effects of extinction are most prominent, the dust parameters are determined less accurately in the monochromatic case compared to the oligochromatic fit. This can be explained when we keep in mind the problems that the monochromatic fitting procedures have to deal with. As discussed in \cite{2013A&A...550A..74D} monochromatic fits to edge-on spiral galaxies experience some degeneracy when trying to determine the dust disc. A dense, small dust disc or a less opaque but radially more extended dust disc can both give very similar images when looking at one wavelength. When combining information from blue and red bands, however, we can partially lift this degeneracy as in this case a denser dust disc would affect the red bands more heavily. Consequently a fitting procedure incorporating this information would be able to differentiate these two cases and yield better constraints on the dust disc parameters.

\section{Fitting results}
\label{Results.sec}

\begin{table*}
\centering
\begin{tabular}{ccr@{$\,\pm\,$}lr@{$\,\pm\,$}lr@{$\,\pm\,$}lr@{$\,\pm\,$}lr@{$\,\pm\,$}lr@{$\,\pm\,$}l}
\hline\hline\\[-0.5ex]
Parameter & unit & 
\multicolumn{2}{c}{IC\,2098} & 
\multicolumn{2}{c}{UGC\,4136} & 
\multicolumn{2}{c}{IC\,2461} & 
\multicolumn{2}{c}{UGC\,5481} & 
\multicolumn{2}{c}{NGC\,3650} & 
\multicolumn{2}{c}{NGC\,3987} \\[2ex]
 \hline \\[-0.5ex]
$h_{R,*}$ & kpc & 4.71 & 0.06 & 6.41 & 0.20 & 3.42 & 0.09 & 4.24 & 0.10 & 3.66 & 0.08 & 7.36 & 0.28 \\
$h_{z,*}$ & kpc & 0.43 & 0.06 & 0.66 & 0.04 & 0.15 & 0.01 & 0.59 & 0.06 & 0.62 & 0.02 & 1.08 & 0.05 \\
$R_{\text{eff}}$ & kpc & 2.88 & 0.54 & 1.19 & 0.07 & 2.83 & 0.26 & 5.99 & 0.59 & 0.91 & 0.09 & 1.79 & 0.12 \\
$n$ & --- & 3.0 & 0.7 & 2.0 & 0.2 & 4.0 & 0.1 & 6.1 & 0.4 & 2.9 & 0.7 & 1.2 & 0.2 \\
$q$ & --- & 0.35 & 0.03 & 0.85 & 0.01 & 0.66 & 0.04 & 0.39 & 0.01 & 0.83 & 0.03 & 0.48 & 0.01 \\
$h_{R,\text{d}}$ & kpc & 2.6 & 0.18 & 29.33 & 0.46 & 5.65 & 1.47 & 10.33 & 1.51 & 4.58 & 1.26 & 4.62 & 0.39 \\
$h_{z,\text{d}}$ & kpc & 0.20 & 0.02 & 0.33 & 0.04 & 0.12 & 0.03 & 0.51 & 0.09 & 0.15 & 0.01 & 0.40 & 0.02  \\
$M_{\text{d}}$ & $10^7~M_\odot$ & 1.3 & 0.1 & 22.0 & 1.6 & 2.3 & 0.8 & 15.2 & 3.0 & 1.4 & 0.3 & 6.1 & 0.5 \\
$\tau_{\text{V}}^{\text{f}}$ & --- & 1.65 & 0.17 & 0.24 & 0.10 & 0.62 & 0.11 & 1.22 & 0.32 & 0.57 & 0.17 & 1.98 & 0.17  \\
$\tau_{\text{V}}^{\text{e}}$ & --- & 21.9 & 1.2 & 17.5 & 1.5 & 29.8 & 4.9 & 25.0 & 8.7 & 17.4 & 2.1 & 23.7 & 1.2  \\
$i$ & deg & 88.9 & 0.1 & 85.5 & 0.1 & 87.1 & 0.2 & 82.3 & 0.6 & 89.0 & 0.2 & 89.9 & 0.1  \\
$L_{g}^{\text{tot}}$ & $10^9~L_{\odot}$ & 0.75 & 0.12 & 4.12 & 0.22 & 2.59 & 0.48 & 8.71 & 1.43 & 1.77 & 0.26 & 1.74 & 0.23 \\ 
$L_{r}^{\text{tot}}$ & $10^9~L_{\odot}$  & 1.20 & 0.15 & 6.69 & 0.42 & 3.74 & 0.51 & 12.93 & 1.49 & 2.83 & 0.39 & 3.12 & 0.36 \\ 
$L_{i}^{\text{tot}}$ & $10^9~L_{\odot}$  & 1.97 & 0.23 & 8.92 & 0.54 & 4.74 & 0.45 & 16.65 & 1.76 & 3.84 & 0.52 & 4.47 & 0.46 \\ 
$L_{z}^{\text{tot}}$ & $10^9~L_{\odot}$  & 2.85 & 0.05 & 11.79 & 0.31 & 6.43 & 0.32 & 21.6 & 0.57 & 4.99 & 0.06 & 6.3 & 0.11 \\ 
$B/T_{g}$ & --- & 0.33 & 0.06 & 0.45 & 0.03 & 0.47 & 0.1 & 0.61 & 0.12 & 0.30 & 0.05 & 0.41 & 0.06 \\ 
$B/T_{r}$ & --- & 0.42 & 0.06 & 0.48 & 0.04 & 0.53 & 0.09 & 0.64 & 0.09 & 0.32 & 0.04 & 0.45 & 0.06 \\ 
$B/T_{i}$ & --- & 0.46 & 0.02 & 0.47 & 0.01 & 0.56& 0.02 & 0.64 & 0.02 & 0.32 & 0.01 & 0.47 & 0.01 \\ 
$B/T_{z}$ & --- & 0.48 & 0.02 & 0.48 & 0.01 & 0.55 & 0.03 & 0.59 & 0.01 & 0.33 & 0.01 & 0.48 & 0.01 \\ [2ex]
\hline\hline\\[-0.5ex]
Parameter & unit & 
\multicolumn{2}{c}{NGC\,4175} & 
\multicolumn{2}{c}{IC\,3203} & 
\multicolumn{2}{c}{IC\,4225} & 
\multicolumn{2}{c}{NGC\,5166} & 
\multicolumn{2}{c}{NGC\,5908} & 
\multicolumn{2}{c}{UGC\,12518} \\[2ex]
\hline \\[-0.5ex]
$h_{R,*}$ & kpc & 2.78 & 0.12 & 5.56 & 0.09 & 3.38 & 0.05 & 3.92 & 0.17 & 4.71 & 0.28 & 2.92 & 0.20  \\
$h_{z,*}$ & kpc &  0.33 & 0.04 & 0.83 & 0.02 & 0.84 & 0.04 & 0.66 & 0.04 & 0.44 & 0.02 & 0.29 & 0.05   \\
$R_{\text{eff}}$ & kpc & 0.92 & 0.05 & 1.83 & 0.06 & 1.25 & 0.04 & 3.17 & 0.59 & 7.54 & 0.42 & 4.81 & 1.16 \\
$n$ & --- & 0.9 & 0.1 & 1.2 & 0.1 & 1.1 & 0.1 & 5.8 & 0.9 & 5.4 & 0.8 & 5.2 & 1.2 \\
$q$ & --- & 0.75 & 0.04 & 0.73 & 0.03 & 0.69 & 0.03 & 0.26 & 0.04 & 0.63 & 0.02 & 0.50 & 0.04 \\
$h_{R,\text{d}}$ & kpc & 3.33 & 0.46 & 11.98 & 2.81 & 10.01 & 2.74 & 5.81 & 1.12 & 5.74 & 1.87 & 8.51 & 1.95 \\
$h_{z,\text{d}}$ & kpc & 0.25 & 0.07 & 0.22 & 0.02 & 0.24 & 0.02 & 0.34 & 0.06 & 0.12 & 0.04 & 0.14 & 0.02 \\
$M_{\text{d}}$ & $10^7~M_\odot$ & 1.5 & 0.4 & 3.4 & 0.7 & 2.1 & 0.5 & 4.8 & 0.9 & 10.8 & 1.0 & 3.2 & 1.6 \\
$\tau_{\text{V}}^{\text{f}}$ & --- & 1.17 & 0.19 & 0.20 & 0.08 & 0.18 & 0.17 & 1.23 & 0.37 & 2.84 & 0.99 & 0.90 & 0.25  \\
$\tau_{\text{V}}^{\text{e}}$ & --- & 15.5 & 2.2 & 11.0 & 1.6 & 7.4 & 0.8 & 21.2 & 4.7 & 132 & 56 & 19.6 & 2.9 \\
$i$ & deg & 83.2 & 0.4 & 87.6 & 0.4 & 89.3 & 0.1 & 87.6 & 0.5 & 83.4 & 0.4 & 87.3 & 0.1  \\
$L_{g}^{\text{tot}}$ & $10^9~L_{\odot}$ & 1.24 & 0.19 & 3.03 & 0.24 & 1.63 & 0.24 & 4.57 & 0.91 & 5.69 & 0.35 & 0.95 & 0.14 \\
$L_{r}^{\text{tot}}$ & $10^9~L_{\odot}$ & 1.86 & 0.23 & 5.07 & 0.37 & 2.48 & 0.26 & 7.18 & 1.14 & 10.91 & 0.61 & 1.67 & 0.15 \\
$L_{i}^{\text{tot}}$ & $10^9~L_{\odot}$ & 2.58 & 0.3 & 6.87 & 0.47 & 3.27 & 0.28 & 9.58 & 1.16 & 16.08 & 0.91 & 2.44 & 0.23 \\
$L_{z}^{\text{tot}}$ & $10^9~L_{\odot}$ & 3.34 & 0.17 & 9.26 & 0.1 & 4.16 & 0.1 & 11.39 & 0.51 & 22.09 & 0.5 & 3.17 & 0.1 \\ 
$B/T_{g}$ & --- & 0.26 & 0.05 & 0.34 & 0.03 & 0.39 & 0.06 & 0.45 & 0.1 & 0.55 & 0.04 & 0.52 & 0.09 \\
$B/T_{r}$ & --- & 0.32 & 0.05 & 0.37 & 0.03 & 0.4 & 0.05 & 0.49 & 0.09 & 0.57 & 0.04 & 0.53 & 0.06 \\
$B/T_{i}$ & --- & 0.34 & 0.02 & 0.38 & 0.01 & 0.4 & 0.01 & 0.49 & 0.02 & 0.58 & 0.01 & 0.54 & 0.03 \\ 
$B/T_{z}$ & --- & 0.39 & 0.02 & 0.4 & 0.01 & 0.43 & 0.01 & 0.45 & 0.02 & 0.55 & 0.01 & 0.52 & 0.02 \\[2ex]

 \hline\hline
 \end{tabular}
 \caption{Results of the oligochromatic FitSKIRT radiative transfer fits to the 12 edge-on spiral galaxies in our sample. For each galaxy we list all physical free parameters with their 1$\sigma$ error bar as determined by FitSKIRT. For the meaning of the different parameters and determination of the error bars, see Sect.~{\ref{Model.sec}} of this paper and/or Sect.~3.1 from \citet{2013A&A...550A..74D}.}
\label{Results.tab}
\end{table*}

We have applied FitSKIRT to the set of 12 edge-on spiral galaxies selected from the CALIFA survey. The computations were done using the high performance cluster of the Flemish Supercomputer Center (Vlaams Supercomputer Centrum, VSC), located at the Ghent University premises. As described in Sect. \ref{SampleSelection.sec}, we have used the SDSS $g$, $r$, $i$ and $z$-band images for each galaxy, which are presented in the left-hand panels of Figure~{\ref{Fits.fig}}. The central column contains the fitted images in the corresponding bands, and the right-hand column gives the residual images. The values of free parameters in the model for the 12 galaxies, with their error bars, can be found in Table~{\ref{Results.tab}}. 

\begin{figure*}
\centering
\includegraphics[width=\textwidth]{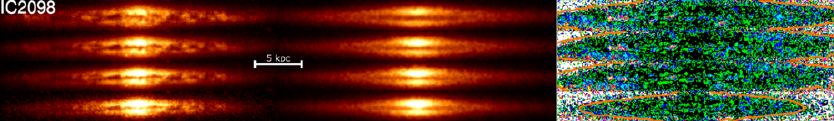}\\[1ex]
\includegraphics[width=\textwidth]{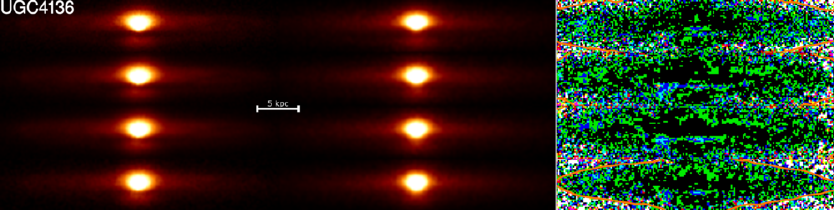}\\[1ex]
\includegraphics[width=\textwidth]{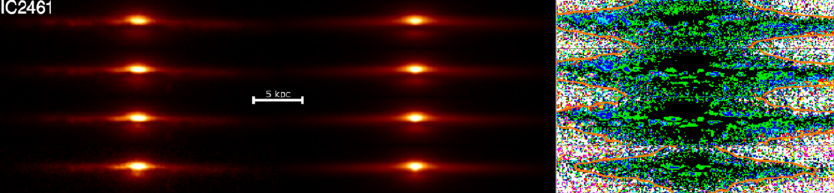}\\[1ex]
\includegraphics[width=\textwidth]{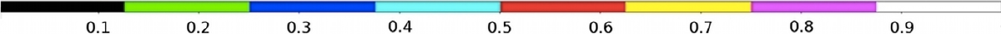}
\caption{Results of the oligochromatic FitSKIRT radiative transfer fits to the 12 edge-on spiral galaxies in our sample. In each panel, the left-hand column represents the observed images in the $g$, $r$, $i$ and $z$ bands (from top to bottom), and the middle column contains the corresponding fits in the same bands. The right-hand column contains the residual images, that indicate the relative deviation between the fit and the image. A color bar with the scaling of the latter is indicated at the bottom. The orange contour in the residuals frames represents the signal-to-noise $3$ level of the reference image. }
\label{Fits.fig}
\end{figure*}

\addtocounter{figure}{-1}
\begin{figure*}
\centering
\includegraphics[width=\textwidth]{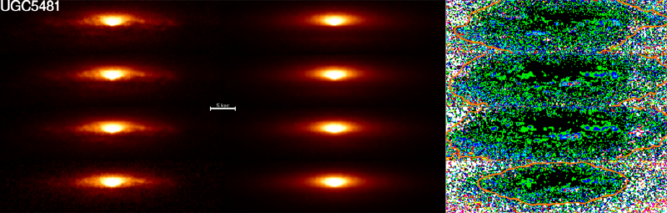}\\[1ex]
\includegraphics[width=\textwidth]{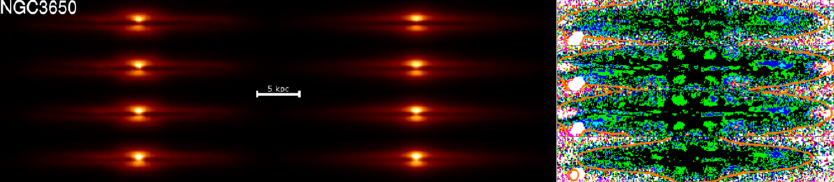}\\[1ex]
\includegraphics[width=\textwidth]{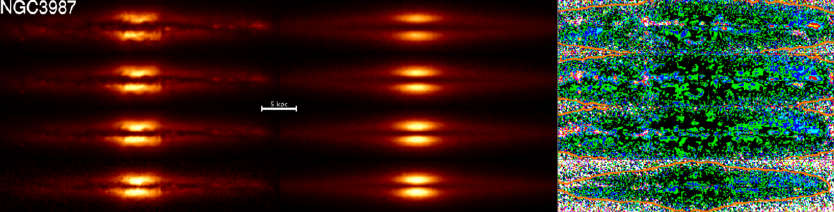}\\[1ex]
\includegraphics[width=\textwidth]{ColorBar.pdf}
\caption{(continued)}
\end{figure*}

\addtocounter{figure}{-1}
\begin{figure*}
\centering
\includegraphics[width=\textwidth]{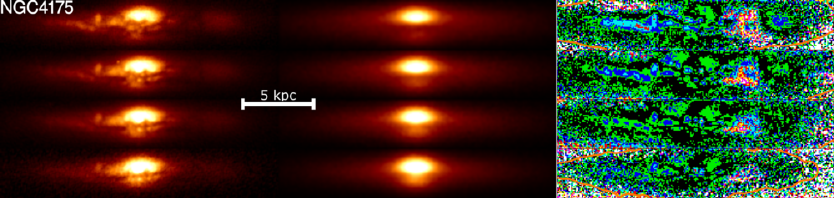}\\[1ex]
\includegraphics[width=\textwidth]{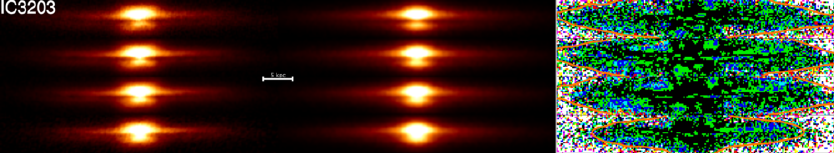}\\[1ex]
\includegraphics[width=\textwidth]{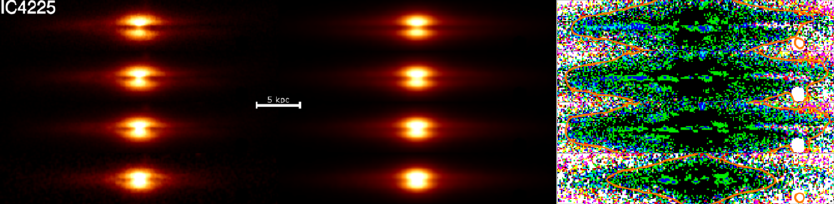}\\[1ex]
\includegraphics[width=\textwidth]{ColorBar.pdf}
\caption{(continued)}
\end{figure*}

\addtocounter{figure}{-1}
\begin{figure*}
\centering
\includegraphics[width=\textwidth]{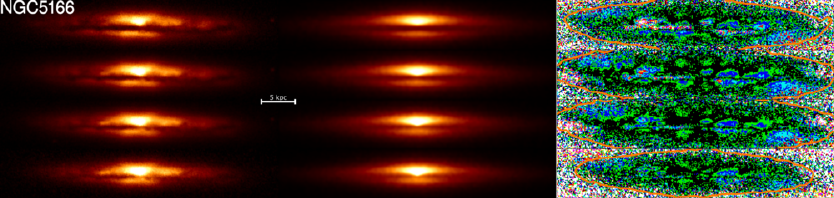}\\[1ex]
\includegraphics[width=\textwidth]{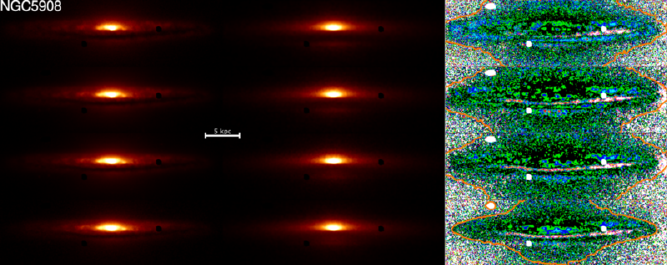}\\[1ex]
\includegraphics[width=\textwidth]{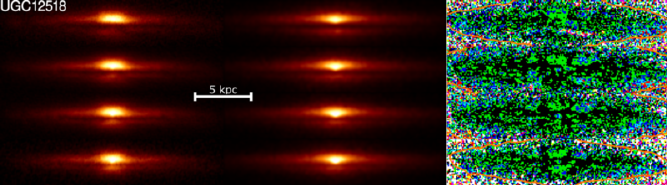}\\[1ex]
\includegraphics[width=\textwidth]{ColorBar.pdf}
\caption{(continued)}
\end{figure*}

\subsection{IC\,2098} 
This galaxy is a late-type Scd, at a distance of about 47.8 Mpc (the mean of the Tully-Fisher based distances determined by \citealt{2007A&A...465...71T} and \citealt{2010Ap&SS.325..163P}). Both the stellar and dust disc of this galaxy have a fairly irregular morphology, which is not unexpected for a late-type galaxy. A warp is not present \citep{2003A&A...399..457S}. In spite of the flocculent morphology, our smooth model fit is quite satisfactory, with the relative difference between observation and model below 25\% over the entire extent of the galaxy. Looking at the parameters in Table~{\ref{Results.tab}}, we note that the dust disc seems to be smaller than the stellar disc in the radial direction. The radial extent of the dust disc might be underestimated by the fitting mechanism since the brighter regions have considerably more weight in our $\chi^2$ objective function. It has to be noticed this is not the same as the degeneracy between a dense dust disc or larger, less dense disc discussed in Sect. \ref{Validation.sec} as in this case all bands would be affected. Consequently the face-on optical depth of this galaxy might be overestimated as the central attenuation is fitted using a higher density disc instead of a more radially extended, optically thinner disc. 

\subsection{UGC\,4136} 
At a distance of 90.8 Mpc \citep{1999PASP..111..438F}, this bulge-dominated Sa-type galaxy \citep{2010ApJS..186..427N} is one of the most distant galaxies in our sample. The best fitting model provides a satisfactory fit to the images in all four bands, although there is a clear signature of the dust lanes visible in the residual maps, especially in the $r$ and $i$ bands. The radial scale length of the dust disc on the other hand is extremely large, especially compared to the moderate stellar scale length. Actually, the dust scale length was found to tend towards the maximum value allowed in the fitting procedure, even when this value was increased. The most likely explanation is that this galaxy has an intrinsic dust distribution which is not properly described by our analytic model. The dust distribution may have a ring structure rather than an exponential disc, which is not uncommon for early-type disc galaxies \citep[e.g.,][]{2011PASP..123.1347K, 2012A&A...543A.161C, 2012MNRAS.419..895D}. When we try to fit a galaxy with a central dust cavity using an exponential disc model, it is natural that the disc scale length will tend towards extremely large values (i.e.\ flat distributions), as this is the only way to avoid a strong central concentration. As the dust distribution can not be described accurately with the use of an exponential disc, this results found for this galaxy were not included in the discussion of a sample as a whole.

\subsection{IC\,2461} 
This galaxy, with a classification varying from S0 \citep{2003PASP..115.1280V} to Sbc \citep{2008A&A...488..523H}, has received quite some attention in the literature due to the fact that it was the host galaxy of the type II supernova SN\,2002bx \citep{2002IAUC.7868....2M}. Fairly recently, it was detected from soft to ultra-hard X-rays \citep{2009ApJ...705..454N, 2010A&A...510A..48C, 2011ApJ...739...57K}, revealing a deeply buried Seyfert 2 nucleus \citep{2012A&A...545A.101P}. In the optical SDSS images, it is a regular galaxy without evidence for perturbations. We assumed a distance of 54.2 Mpc ( the average of Tully-Fisher based distances \citep{2007A&A...465...71T, 2009ApJS..182..474S, 2010Ap&SS.325..163P}). The FitSKIRT model reproduces the images quite accurately. Not only do the residuals show very little deviation or irregularities, also the error bars on each of the derived parameters are modest.

\subsection{UGC\,5481}  
This rather anonymous galaxy at a distance of 90.2 Mpc \citep{2007A&A...465...71T} is one of the smallest galaxies in the sample in angular size. With an axis ratio of over 25\% it is the galaxy furthest from edge-on (the fitted value of the inclination is $82.3\pm0.6$ deg). As a result, the dust lane is not as sharp and prominent as for most other galaxies in the sample. In spite of this rather large deviation from edge-on, the fits are surprisingly satisfactory. The error bars on the fitted parameters are not larger than average, and the residual maps are quite smooth and without strong irregularities. 

\subsection{NGC\,3650} 
This nice Sb type galaxy at a distance of 59.7 Mpc \citep{1999PASP..111..438F} is very close to exactly edge-on and characterized by a very clear and strong dust lane. According to \citet{1990MNRAS.246..458S}, it has a barely perceptible warp. The bulge is strongly peanut-shaped \citep{2000A&AS..145..405L}, which shows as a X-shape pattern in the residual maps (our standard model of a flattened S\'ersic bulge cannot reproduce peanut-shaped bulges). Apart from this feature, the FitSKIRT fits reproduce the SDSS images particularly well and the parameters of the model are well constrained.  

\subsection{NGC\,3987} 
This almost exactly edge-on Sb galaxy ($i=89.9\pm0.2$ deg) shows a very prominent, albeit somewhat irregularly shaped dust lane. It was one of the first edge-on galaxies in which the surface brightness distribution was studied in detail \citep{1975AJ.....80..188D}. It is one of the best known galaxies in our sample, mainly thanks to hosting the type Ia supernova SN\,2001V \citep{2003A&A...397..115V}. Redshift-independent distances in the literature vary from 41 Mpc \citep{2010ApJ...716..712A} to 70 Mpc \citep{2007A&A...465...71T}; we use an average value 61.4 Mpc. NGC\,3987 is a bright radio source and strong emitter at infrared and sub-millimeter wavelengths \citep{2001ApJ...554..803Y, 2005MNRAS.364.1253V}. The FitSKIRT fit to the SDSS images is not perfect, with many artifacts visible in the residual maps. Looking at the best fitting parameters, we note that the stellar disc seems to be more radially extended than the dust disc, as we previously obtained for IC\,2098. In this case, however, the dust disc is probably underestimated due to a warp in the dust lane. Indeed, the residual frames clearly show the region where the fit and model differ significantly: the dust lane deviates from the major axis in opposite directions on either side of the galaxy. Interestingly, \citet{2005MNRAS.364.1253V} have observed that the 850 $\mu$m submillimeter emission in NGC\,3987 is slightly offset from the galaxy's plane. As our analytic model does not include such feature it will break off at radius where the dust disc deviates away from the plane of the galaxy. The resulting face-on optical depth is therefore probably overestimated. 

\subsection{NGC\,4175} 
Together with the two lenticular galaxies NGC\,4169 and NGC\,4174, this Sbc type galaxy forms the \citet{1982ApJ...255..382H} compact group HCG\,61. At a distance of about 42.2 Mpc \citep{2007A&A...465...71T}, NGC\,4175 is the most nearby galaxy in our sample. It has an active galactic nucleus \citep{2010AJ....139.1199M} with a strong emission at MIR, FIR and radio wavelengths \citep[e.g.,][]{2004AJ....127.3235S, 2007AJ....134.1522J, 2008ApJ...673..730G}, and a significantly disturbed morphology in the inner regions. The SDSS images show a clear asymmetry between the two sides of the disc in all four bands. FitSKIRT is unable to fit these asymmetries using our smooth 2D model, so these features stand out in the residual maps as well. Nevertheless the fit is acceptable when looking at the residuals and most parameters are quite well constrained. 

\subsection{IC\,3203} 
At a distance of 119.2 Mpc \citep{1997ApJS..109..333W, 2007A&A...465...71T}, IC\,3203 is the most distant galaxy from the sample, and among the smallest on the sky. It has been classified both as S0 \citep{2003PASP..115.1280V} and Sb \citep{2008A&A...488..523H}, and was home of a rare type IIb supernova, SN\,2003ac \citep{2003IAUC.8085....2F}. As it is resolved quite poorly, the uncertainty on the dust parameters is larger compared to other galaxies. The residual maps indicate that the bulge is moderately peanut-shaped. The face-on optical depth derived from the FitSKIRT fit ($\tau_V^{\text{f}} = 0.20\pm0.08$) is quite low and might be underestimated because of the poorly resolved and smoothened dust lane.

\subsection{IC\,4225} 
IC\,4225 is a rather anonymous Sa galaxy in the Coma supercluster region at a distance of about 73.9 Mpc \citep{1995AJ....109.1458R}. It is oriented almost perfectly edge-on ($i=89.3\pm0.1$ deg), but unfortunately has the smallest angular extent of all galaxies in our sample. The SDSS images and the residual images clearly show an asymmetry between the left an right side of the galaxy, which might be due to spiral structure. In general, however, the FitSKIRT fits are satisfactory with residual errors below 25\% for most of the pixels in the images. 

\subsection{NGC\,5166} 
NGC\,5166 is a rather ordinary Sb type galaxy at a Tully-Fisher based distance of 62.6 Mpc \citep{2007A&A...465...71T}. It has a prominent but somewhat asymmetric dust lane. Its bulge is known to be strongly peanut-shaped \citep{2000A&AS..145..405L}, which shows up in the residual frames (in this case mostly in the $r$-band) as the typical butterfly or X-shape pattern. Apart from these features, the overall radiative transfer fit is remarkably good.

\subsection{NGC\,5908} 
This Sb galaxy forms a weakly interacting pair with NGC\,5905 at a distance of 53.4 Mpc \citep{2007A&A...465...71T}. Its surface brightness distribution has been studied extensively in the past two decades \citep[e.g.][]{1997NewA....1..349P, 2000A&AS..141..103L, 2000A&AS..144...85S}. Similarly to UGC\,5481 and NGC\,4175, it has a fairly large deviation from an edge-on inclination ($i\sim83$ deg). In apparent size, it is the largest galaxy in our sample, but the poorest radiative transfer fit: the fitted edge-on optical depth value is unrealistically high at $\tau_V^{\text{e}}\sim132$, and strong structures remain visible in the residual images. Although the bright regions are recovered reasonably, the dust lane is fitted very poorly with relative errors exceeding 100\%. Both the observations and the residuals show that the dust is distributed in a smooth ring rather than the standard double-exponential disc we have used for our fits. In fact, with its relatively smooth dust ring, bright bulge and extended halo, the galaxy reminds strongly of the Sombrero galaxy \citep{2012MNRAS.423..877G, 2012MNRAS.419..895D}, a fact already noted by \citet{1964rcbg.book.....D}. Because the geometry clearly deviates from an exponential disc, this galaxy was discarded when determining the average quantities of our sample.

\subsection{UGC\,12518} 
The last galaxy in our sample is again a rather unexceptional Sb galaxy, located at 51.6 Mpc \citep{2009ApJS..182..474S}. Together with IC\,3203, it has the smallest apparent vertical size of the entire sample. Consequently, the dust lane is only a few pixels thick making it hard to get accurate constraints on the dust distribution parameters. Moreover, the SDSS images and the residual images show that the bulge is clearly box-shaped. The result is that the fit is moderately good, although the dust parameters are recovered with relatively large error bars.

\section{Discussion}
\label{Discussion.sec}
\subsection{Quality of the FitSKIRT radiative transfer fits}

\begin{figure}
\centering
\includegraphics[width=0.45\textwidth]{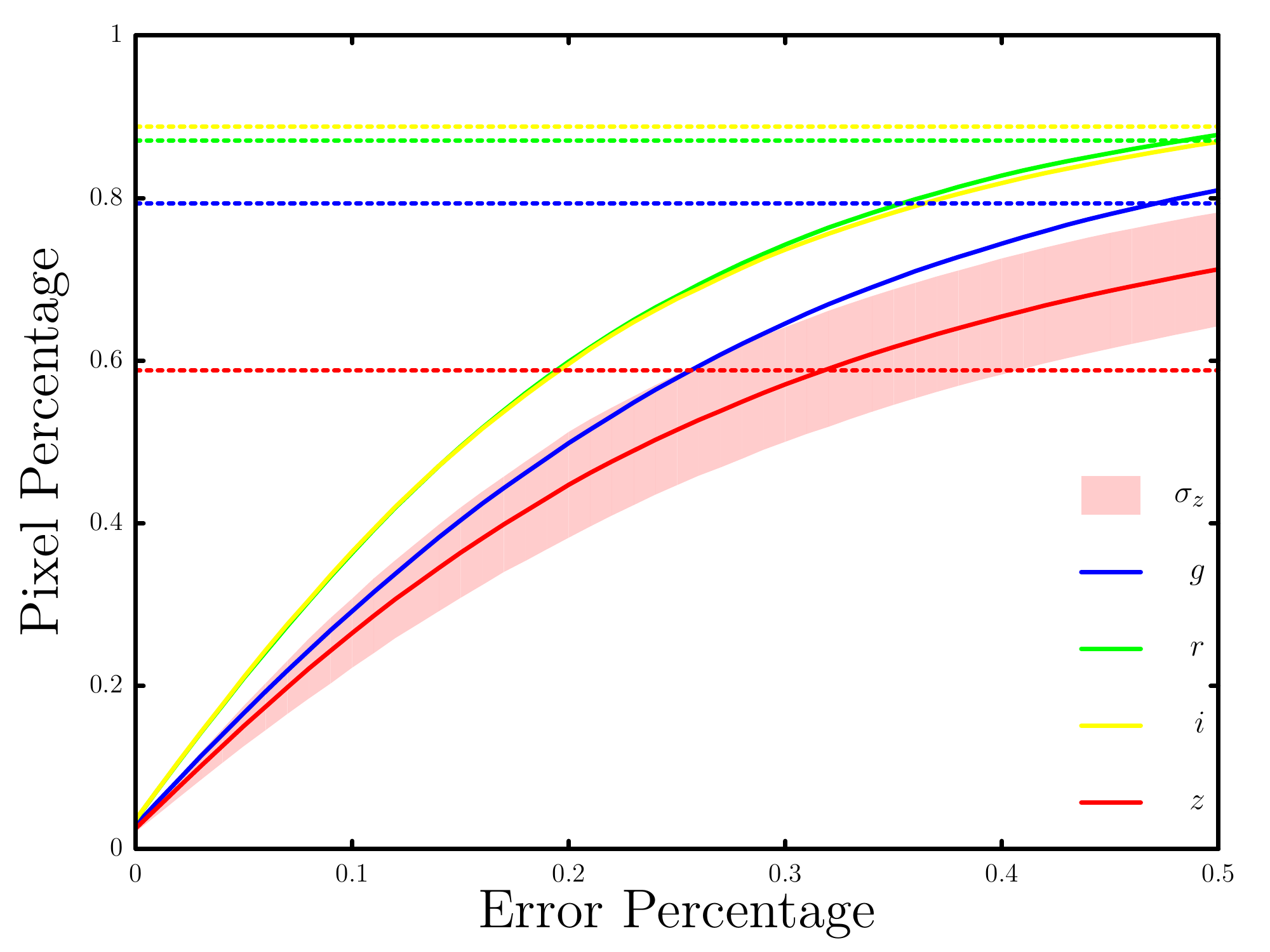}
\caption{Cumulative pixel distribution of all the residual frames for all 12 galaxies. The standard deviation within the sample is given for the z-band. The dashed lines represent the fraction of the pixels in the observed images with a signal-to-noise of three or higher.}
\label{Residuals.fig}
\end{figure}

Looking at the results, we conclude that in general most galaxies are modeled accurately, especially when keeping in mind that the FitSKIRT models only consist of three basic components and they were determined by an automated procedure over a large parameters space without strong initial boundary conditions. Only two of our galaxies, NGC\,5980 and UGC\,4136, could not be accurately described by the model we have presumed. The dust geometry in these two systems is probably better described by a ring-like geometry rather than an exponential disc. The values found by FitSKIRT do not have any significant meaning for these two galaxies and using them could bias our results. Although it is likely that other galaxies in our sample also have a geometry that deviates from our model, we did not find strong indications of systematic deviations. Most of the variations seen in the residual maps are the result of small perturbations in the galaxy rather than a larger-scale and systematic deviation.

To quantify the accuracy of the fits, Figure~{\ref{Residuals.fig}} shows the cumulative pixel distribution of the residuals of the entire sample, averaged for each band separately. The solid lines on this figure show which fraction of the pixels are reproduced by the models with a relative error smaller than the value on the horizontal axis, the shaded region indicates the dispersion on this curve for the $z$-band (it is representative for the other bands as well). The dashed lines represent the fraction of the pixels in the observed images with a signal-to-noise of three or higher. 

The $r$ and $i$-band images show the least deviation between the model and the corresponding SDSS reference images. In other words, FitSKIRT provides the best fits to the $r$ and $i$ band images. The $g$-band images are fitted slightly worse -- the reason is probably that the dust extinction is strongest in this band and that asymmetries and small-scale irregularities are most prominent. Finally, the $z$-band observations, where only about $60\%$ of the pixels have a signal-to-noise above three, are fitted least accurately. The most likely reason is the relatively poor signal-to-noise ratio. In Sect. \ref{Validation.sec} it was shown that in our test case, where the signal-to-noise was equal in all bands, there was no significant biasing of one band over the other. 
In general, we find that, in any band, more than half of the pixels have a deviation of $25\%$ or better, and this improves to 15\% in the $r$ and $i$ bands. 

\begin{table}
\centering
\begin{tabular}{ccr@{$\,\pm\,$}lc}
\hline\hline\\[-0.5ex]
Parameter & unit & mean & RMS & 1$\sigma$ (\%) \\[2ex]
\hline \\[-0.5ex]
$h_{R,*}$ & kpc & 4.23 & 1.23 & 3 \\
$h_{z,*}$ & kpc &  0.51 & 0.27 & 7 \\
$R_{\text{eff}}$ & kpc & 2.31 & 1.59 & 11 \\
$n$ & --- & 2.61 & 1.80 & 14 \\
$q$ & --- & 0.56 & 0.20 & 6 \\
$h_{R,\text{d}}$ & kpc & 6.03 & 2.92 & 19 \\
$h_{z,\text{d}}$ & kpc & 0.23 & 0.10 & 13 \\
$M_{\text{d}}$ & $10^7~M_\odot$ & 3.02 & 2.21 & 23 \\
$\tau_{\text{V}}^{\text{f}}$ & --- & 0.76 & 0.60 & 28 \\
$\tau_{\text{V}}^{\text{e}}$ & --- & 18.0 & 7.1 & 15 \\
$i$ & deg & 86.7 & 2.5 & 0.4 \\
$L_{g}^{\text{tot}}$ & $10^9~L_{\odot}$ & 2.70 & 2.40 & 15\\
$L_{r}^{\text{tot}}$ & $10^9~L_{\odot}$ & 4.12 & 3.54 & 12 \\
$L_{i}^{\text{tot}}$ & $10^9~L_{\odot}$ &  5.64 & 4.50 & 10\\
$L_{z}^{\text{tot}}$ & $10^9~L_{\odot}$ & 7.35 & 5.72 & 3  \\ 
$B/T_{g}$ & --- & 0.41 & 0.11 & 17 \\
$B/T_{r}$ & --- &  0.45 & 0.11 & 14\\
$B/T_{i}$ & --- & 0.46 & 0.11 & 4  \\ 
$B/T_{z}$ & --- &  0.46 & 0.08 & 3 \\[2ex]
 \hline\hline
 \end{tabular}
 \caption{Mean values, RMS values, and average 1$\sigma$ error bars for the different model parameters for the galaxies in our sample. For the parameters of the dust, UGC\,4163 and NGC\,5908 were not included, as these galaxies are poorly described by our standard double exponential disc model.}
\label{MeanValues.tab}
\end{table}

The last column in Table~{\ref{MeanValues.tab}} lists the average 1$\sigma$ error bar on each of the fitted values for the free parameters in the model. Notice that these errors do not take into account possible systematic errors. The inclination is the best constrained parameter: in all cases it is determined with an accuracy of less than half a degree. The parameters linked to the stellar geometry are better constrained than the dust parameters, which is consistent with the results found in \cite{2013A&A...550A..74D}. The stellar disc parameters are determined with a relative accuracy of less than 10\% while the bulge parameters are derived with an uncertainty of approximately 15\%. The error bars for the scale height of the dust disc and the edge-on optical depth, two parameters that have a direct effect on the appearance of the dust lane in edge-on spiral galaxies, are also of the order of 15\%. For the dust mass and the face-on optical depth, parameters that do not have as direct an impact on the morphology, the error bar is between 20 and 30\%. Both the disc and bulge luminosity are recovered more precisely when we go to redder bands. This is not so peculiar as the attenuation in the blue bands makes it harder to recover the intrinsic luminosity. Overall the deviations on the luminosity are of order of 15\% or less. The third column in Table~{\ref{MeanValues.tab} contains the mean values and the RMS of the physical free parameters derived from the radiative transfer fits.

The general conclusion drawn from our radiative transfer fits is that most galaxies can be modeled appropriately with an exponential disc for the stars and the dust and a S\'ersic profile describing the central bulge. In most cases the parameters of the different components can be constrained with a satisfactory level of accuracy. 

\subsection{The stellar disc and bulge}

As none of the galaxies in our sample have been modeled using radiative transfer simulations before, we cannot directly compare the values we obtain for individual galaxies. As a useful sanity check, however, we can check whether the average values of the derived parameters are in agreement with general values for the global galaxy population. This applies in particular to the parameters of the stellar geometry, which can be derived from large samples of galaxies without the need of full radiative transfer calculations. 

Concerning the properties of the stellar disc, an interesting sample to compare our results to is the set of 34 edge-on spiral galaxies by \citet{2002MNRAS.334..646K}. They fit a combination of a double-exponential disc and a S\'ersic bulge to $I$-band images, carefully masking the dust lane region. Averaging the values of their Table~1, we find a mean scale length of $4.73\pm2.57$~kpc and scale height of $0.57\pm0.25$~kpc. These values are in agreement with our results of $4.23\pm1.23$~kpc and $0.51\pm0.27$~kpc, respectively. The mean intrinsic flattening of the stellar disc, i.e.\ the ratio $h_{R,*}/h_{z,*}$, in the \citet{2002MNRAS.334..646K} sample is $8.21\pm2.36$, again in excellent agreement with our value of $8.26\pm3.44$.

For the properties of the bulge, we can compare our average values to the \citet{2009MNRAS.393.1531G} sample, which contains nearly 1000 face-on galaxies selected from the SDSS for which detailed morphological decomposition was applied. When we eliminate the elliptical galaxies from the sample and look at the $r$-band values, we find for the S\'ersic index a mean value $2.37\pm1.35$, consistent with our value of $2.61\pm1.80$. 

Interesting is the effective radius of the bulge: the typical value obtained by \citet{2009MNRAS.393.1531G} is $0.84\pm0.36$~kpc, which is significantly smaller than (but still compatible within the error bars with) the value of $2.31\pm1.59$~kpc we obtain. One reason for this distinction could be that we have not taken into account a bar in our radiative transfer fits, whereas \citet{2009MNRAS.393.1531G} do fit a combination of a disc, bulge and bar to their images. The addition of a bar in the fitting has a significant effect on the resulting properties of the bulge. A second important aspect is the effect of dust attenuation on the derived properties: the presence of dust can severely affect the apparent parameters in bulge-disc decomposition, even at low inclinations \citep[e.g.][]{2008MNRAS.388.1708G, 2010MNRAS.403.2053G, 2013A&A...553A..80P, 2013A&A...557A.137P}. In particular, \citet{2010MNRAS.403.2053G} demonstrate using detailed radiative transfer modeling that if the effects of dust are not taken into account then bulge effective radii are systematically underestimated. Another possible cause for this difference is that one of the selection criterions for our sample was presence of a dust lane. Galaxies with a large bulge are more likely to show a prominent dust lane than galaxies with a small bulge, so this selection criterion could be in part responsible for the rather large bulges in our sample. 

Similarly interesting is the value of the bulge-to-disc ratio or the bulge-to-total ratio. Since \citet{2009MNRAS.393.1531G} uses a disc, a bar and a bulge in his fits, we can not directly compare the bulge-to-total values between both samples. Consequently we have to add their bulge and bar component when calculating the bulge-to-total ratio which would be found by a two-component fit. This results in $0.32 \pm 0.22$, $0.33 \pm 0.20$ and $0.33 \pm 0.20$ for the $g$-, $r$- and $i$-band respectively. The bulges of the galaxies in our sample seem about 25\% more luminous compared to the sample of \citet{2009MNRAS.393.1531G}. A similar flattening of the bulge-to-total ratio is found when going to longer wavelengths. 

We can conclude that in general the parameters found for the stellar disc and bulge are in good agreement with other studies and there no large systematic deviations in our sample or fitting procedure.

\subsection{The star-dust geometry}                                                                                                                                                                                                                                                                                                                                                                                                                                                                      

\begin{figure*}
\centering
\includegraphics[width=0.47\textwidth]{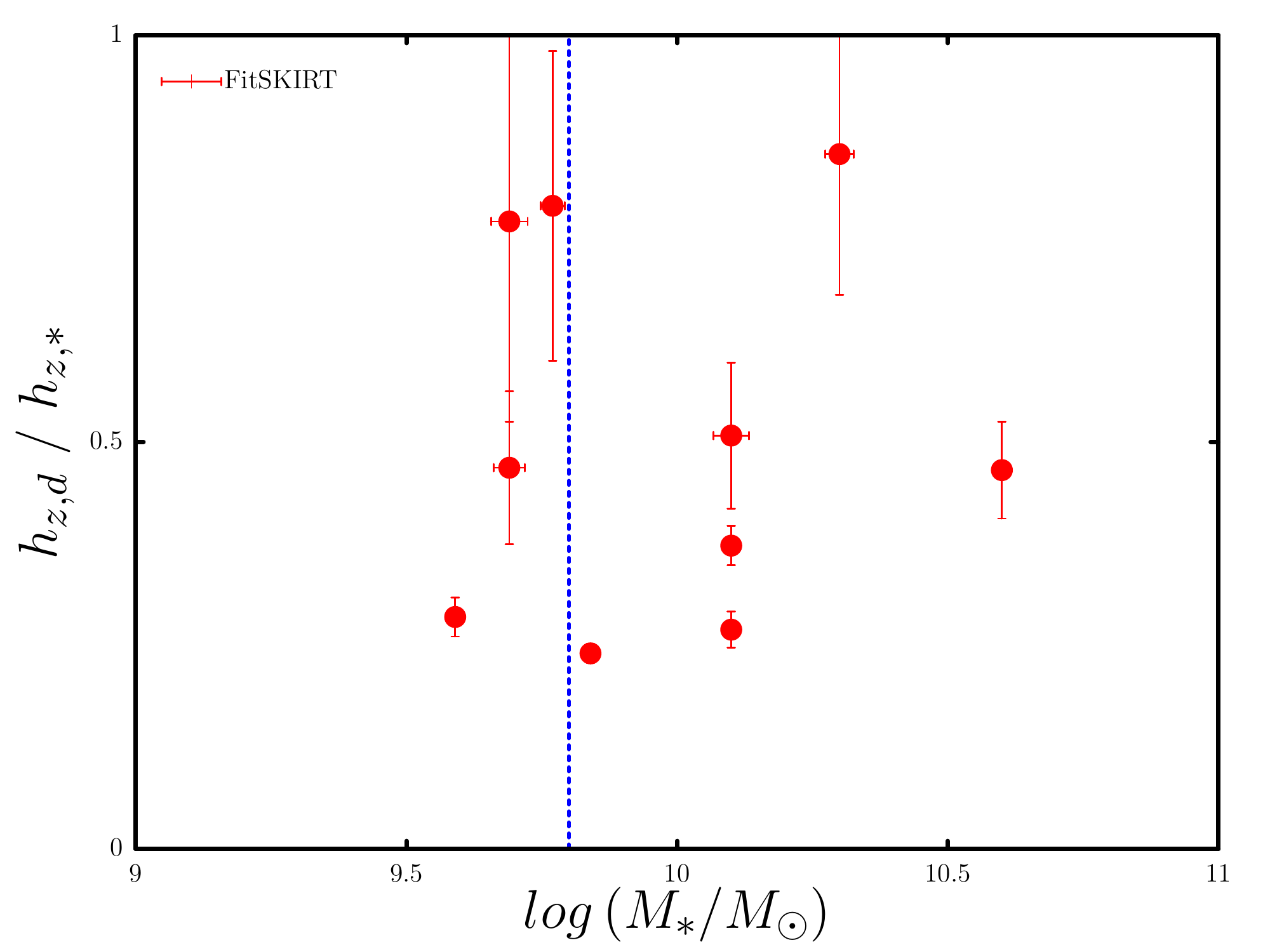}
\includegraphics[width=0.47\textwidth]{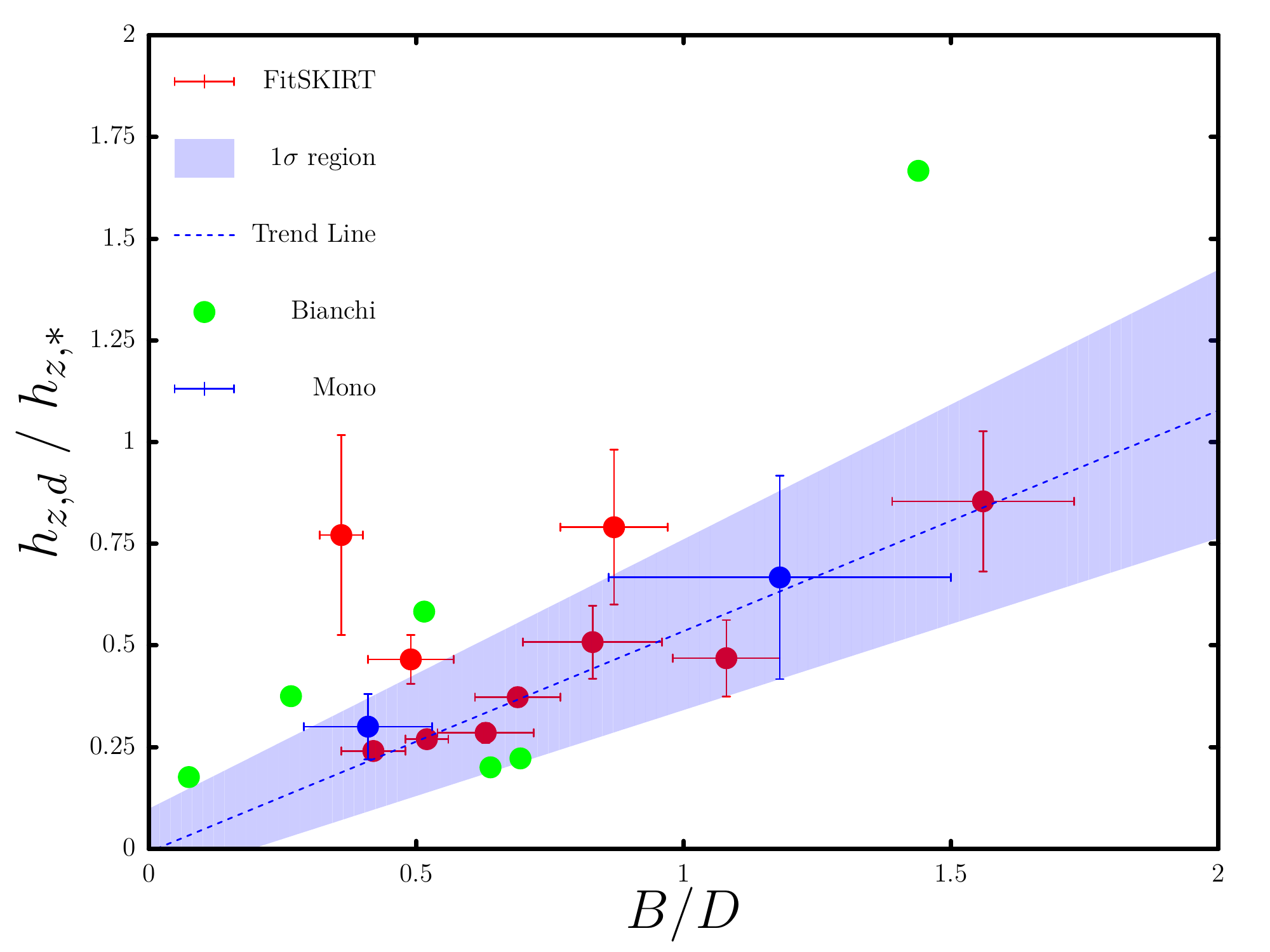}
\caption{Left: The ratio of the dust to stellar scale heights in function of the stellar mass. The blue line marks the stellar mass at the which the rotation speed exceeds 120 km\,s$^{-1}$ marking the transition from dynamically stable discs to perturbed discs prone to vertical "collapse" as discussed in \citet{2004ApJ...608..189D}.
Right: the relation between ratio of the dust scale height and the stellar scale height and the bulge-to-disc ratio for the entire sample. The blue line shows the best fitting linear fit while the blue region is the one-sigma deviation calculated out of the uncertainties on the best fitting values. The blue dots are previous monochromatic FitSKIRT results for NGC\,4565 \citep{2013A&A...550A..74D} and NGC\,4013 \citep{2012MNRAS.427.2797D}. The green dots are the values found by \citet{2007A&A...471..765B}.}
\label{fig:dratios}
\end{figure*}

Table \ref{MeanValues.tab} also lists the average scale length and scale height of the dust disc and their spread within our sample. The dust scale height is in good agreement with the values $0.23 \pm 0.08$ kpc and $0.25 \pm 0.11$ kpc found using radiative transfer fits on V-band images by \citetalias{1999A&A...344..868X} and \citetalias{2007A&A...471..765B} respectively. 

We find a relatively good agreement of the relative sizes of the dust and stellar discs. For the ratios of the dust to stellar scale height we find $0.50 \pm 0.22$, which is consistent with the conclusion found by \citetalias{1999A&A...344..868X}, $0.58 \pm 0.13$, and \citetalias{2007A&A...471..765B}, $0.52 \pm 0.49$. For the dust scale length to stellar scale length ratio we find $1.73 \pm 0.83$. This is slightly larger in average and in spread but still in agreement with \citetalias{1999A&A...344..868X}, $1.36 \pm 0.17$ and \citetalias{2007A&A...471..765B}, $1.53 \pm 0.55$. We can hence confirm the results from these studies, that the dust in spiral galaxies is typically distributed in a disc only half as thick but radially more than $50\%$ larger the stellar disc.

To investigate a similar feature in our sample, we have calculated the stellar masses using the intrinsic luminosities and $g-z$ color difference according to the recipe described in \citet{2009MNRAS.400.1181Z}. The left panel of Figure \ref{fig:dratios} shows the dust to stellar scale heights as a function of stellar mass. The vertical line shows the stellar mass of a bulgeless galaxy corresponding to the rotation speed of 120 km\,s$^{-1}$ using the baryonic Tully-Fisher relation \citep{2001ApJ...550..212B, 2004ApJ...608..189D}. Although  most of our galaxies have a higher stellar mass and are therefore expected to rotate faster, four of the ten galaxies fall below the transition line. It should be noted that using the stellar mass instead of the actual rotation speed is only an approximation and that our sample contains galaxies with moderate to large bulges while the sample of \citet{2004ApJ...608..189D} consists of bulgeless galaxies. 

To check whether the presence of a bulge has an effect on the relative size of the dust disc, we have investigated the dust to stellar scale height in function of the $g$-band bulge-to-disc ratio. These ten galaxies with corresponding uncertainties are shown in the right panel of Figure \ref{fig:dratios}. To extend our sample and as a sanity check we have added two galaxies modeled previously with FitSKIRT in blue dots \citep{2013A&A...550A..74D, 2012MNRAS.427.2797D}, and V-band values of the galaxies modeled by \citetalias{2007A&A...471..765B} in green dots. However, as there was no error determination for the parameters determined by \citetalias{2007A&A...471..765B}, they were not taken into account in the determination of the best fitting trend line. The galaxies modeled by \citetalias{1999A&A...344..868X} do not come with a clear bulge-to-disc ratio and were therefore not included. 

Looking at the right panel of Figure \ref{fig:dratios}, it appears there is a correlation between the ratio of the dust to stellar scale height and the bulge-to-disc luminosity ratios, i.e. galaxies with more prominent bulges tend to have dust scale heights increasingly similar to the stellar scale heights. Again, we have to consider a bias due to our sample selection criteria. As discussed before, galaxies with larger bulges often have more noticeable dust lanes. Therefore it is not surprising to find vertically more extended dust discs in galaxies with a more prominent bulge compared to thinner, bulgeless galaxies. This, however, does not explain why there is a lack of thin dust discs for galaxies with larger bulges. Consequently, this potential bias in the selection process alone can not explain the trend found for this sample.

Another small bias might be the fact that only one exponential disc is used to describe the stellar population. It is known that galaxies can have distinct separation of stellar populations where the young stars are in a thin yet heavily obscured disc while the older stellar population is distributed in a thicker disc \citep[see also][]{2004ApJ...617.1022P, 2011A&A...527A.109P, 2012ApJ...746...70S}. As the thin, bulgeless galaxies typically possess a younger, yet obscured stellar population, the resulting stellar scale height might be overestimated. Therefore the result is a smaller dust to stellar scale height ratio.

Whether or not these biases in the sample can completely explain the correlation is unclear. One possible physical explanation for the lack of thin dust discs for bulge-dominated galaxies may lie in variations of the dominant dust heating mechanisms. Dust has been found to be predominantly heated by either diffuse emission from the total stellar population \citep[e.g.][]{2010A&A...518L..65B, 2012MNRAS.419.1833B, 2011AJ....142..111B, 2012MNRAS.426..892G} or by star-forming regions \citep{2010A&A...518L..55G, 2014arXiv1402.5967H}. In a typical star-forming disk, dust and other ISM material are heated locally by star-forming regions. In galaxies with prominent bulges, the dust heating from star-forming regions may be supplemented with an additional source of dust heating originating from the diffuse emission of the older stellar population in the bulge. One may hypothesize that a greater contribution to the dust heating by the total stellar population in galaxies with increasing bulge-to-disc ratio, may lead to an increase in net heating of the ISM and, subsequently, a thickening of the dust disc. However, whilst the tendency of bulge-dominated galaxies to experience a thickening of the dust disc due to increased dust heating from the stellar population in the bulge may plausibly explain the lack of thin dust discs found at high bulge-to-disc ratios, we strongly warn that this physical mechanism is highly uncertain, speculative, and only included for completeness in our discussion. Further studies, in which we extend our sample size, wavelength range and the capability of the models to describe multiple stellar populations, are necessary in order to provide further insight into a possible physical explanation for this relation.

\subsection{Optical depth}
\label{opticaldepth.sec}

The galaxies analyzed by \citetalias{1999A&A...344..868X}, have a V-band face-on optical depth of $0.49\pm0.19$ and none of them with a value higher than one, which would indicate these galaxies would be completely transparent if they were to be seen face-on. \citetalias{2007A&A...471..765B} used a Monte Carlo radiative transfer code to model a different sample of galaxies (two galaxies are in common with the \citetalias{1999A&A...344..868X} sample). He found a slightly larger spread in V-band optical depths, ranging up to $\tau_{\text{V}}^{\text{f}} = 1.46$ for NGC\,4013. The average value for the sample studied by \citetalias{2007A&A...471..765B} is $0.58\pm0.45$.  

In our sample of 10 galaxies, we find a larger spread in face-on V-band optical depth, ranging from a mere 0.18 for IC\,4225 to 1.98 for NGC\,3987. The average value and RMS for our sample is $0.76\pm0.60$, formally consistent with the mean value found by \citetalias{1999A&A...344..868X} and \citetalias{2007A&A...471..765B}. The larger spread however seems to suggest that spiral galaxies, although on average optically thin, a non-negligible part is not entirely transparent. This slightly larger value for the optical depth can at least partly be ascribed to the different selection criteria. \citetalias{1999A&A...344..868X} and \citetalias{2007A&A...471..765B} considered very nearby galaxies, whereas our galaxies are significantly more distant (the average distances to the galaxies in the samples are 21, 30, and 64 Mpc for \citetalias{1999A&A...344..868X}, \citetalias{2007A&A...471..765B} and our sample respectively). The combination of the larger distance with the requirement that the galaxies need to show a prominent dust lane, might bias our sample toward galaxies with more dust than average, and hence with an average optical depth larger than the more nearby samples of \citetalias{1999A&A...344..868X} and \citetalias{2007A&A...471..765B}. Whether or not this can fully explain the difference is hard to tell. 

One might suspect that differences in the optimization and the radiative transfer treatment itself could be a possible origin for this difference. However, it does not seem likely that FitSKIRT systematically overestimates the optical depth: for NGC\,4013, the one galaxy that has been modeled using the three codes, the optical depth found by FitSKIRT lies in between the values obtained by the two other teams \citep{2013A&A...550A..74D}.

\subsection{Dust mass and the dust energy balance}
\label{masses.sec}

Since the launch of the {\it Herschel} Space Observatory, dust masses for thousands of nearby galaxies have been determined by fitting simple modified blackbody or more complicated SED models to the observed far-infrared and submm fluxes. Our radiative transfer models provide us with a completely independent and alternative technique to estimate the dust mass in galaxies. The average dust mass for our galaxy sample is $\langle \log M_{\text{d}}/M_{\odot}\rangle = 7.48\pm0.32$. When we compare this value to the typical values found for normal late-type galaxies based on {\it Herschel}FIR/submm SEDs, we find fairly compatible results. For example, \citet{2012MNRAS.419.3505D} find $\langle \log M_{\text{d}}/M_{\odot}\rangle = 7.06\pm0.45$ in a 500 $\mu$m selected sample of 78 optically bright galaxies from the HeViCS survey \citep{2010A&A...518L..48D}, \citet{2012MNRAS.425..763G} found $\langle \log M_{\text{d}}/M_{\odot}\rangle = 7.34\pm0.30$ for a sample of 11 nearby galaxies from the KINGFISH sample \citep{2011PASP..123.1347K}, and \citet{2012MNRAS.427..703S} obtained $\langle \log M_{\text{d}}/M_{\odot}\rangle = 8.01$ for a sample of more than thousand $z<0.35$ galaxies selected from the H-ATLAS survey \citep{2010PASP..122..499E}.

\begin{figure}
\centering
\includegraphics[width=\columnwidth]{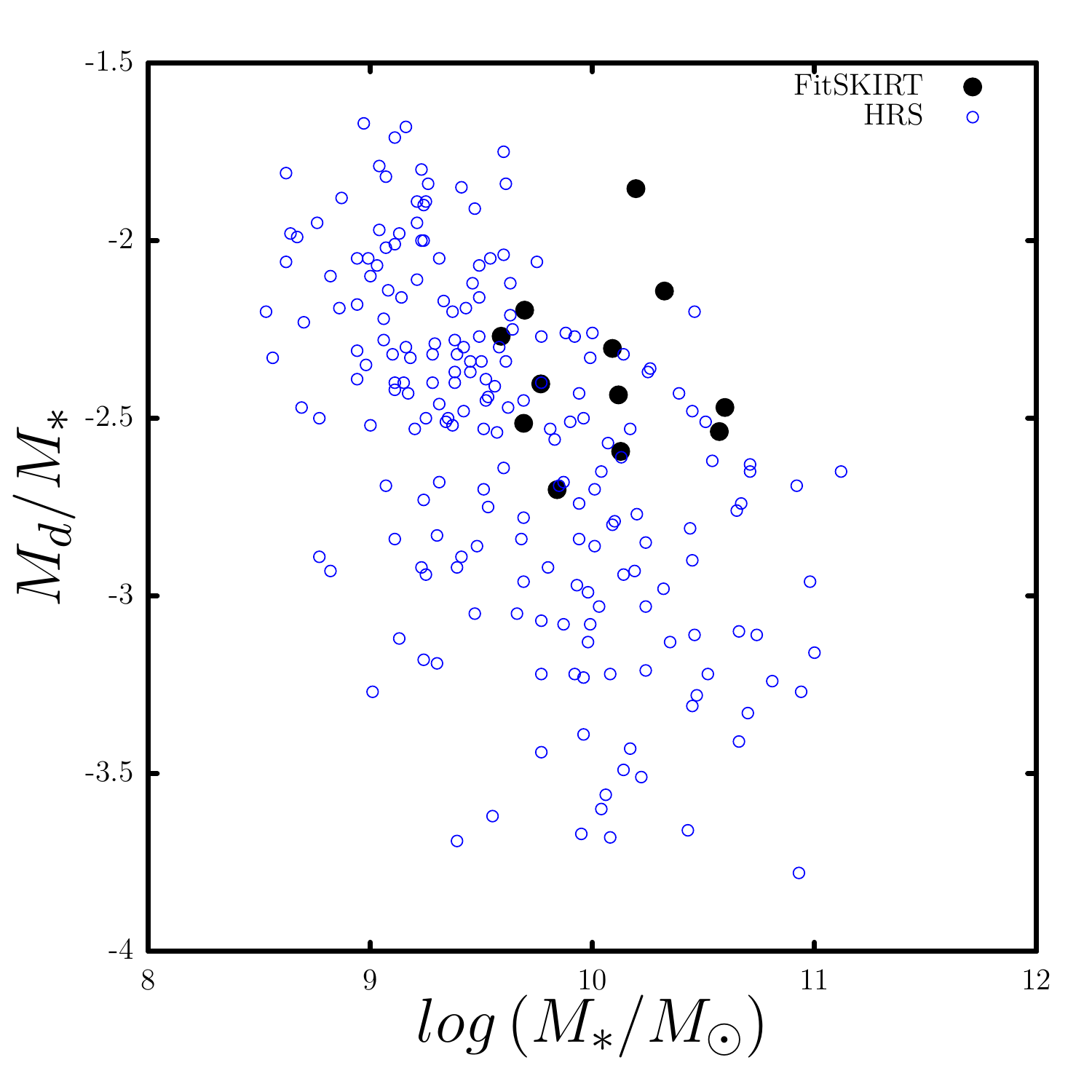}
\caption{The dust to stellar mass ratio (specific dust mass) in function of the stellar mass. The black dots represent the 12 galaxies in our sample while the open circle are the values found for the Herschel Reference Survey galaxies \citep{2012A&A...540A..52C}. Notice that our CALIFA sample seems to be more dust rich compared the HRS sample.}
\label{SpecificDustMass.fig}
\end{figure}

In Figure~{\ref{SpecificDustMass.fig}} we plot the dust-to-stellar mass ratio as a function of the stellar mass for our sample galaxies. We also show on this plot the late-type galaxies from the Herschel Reference Survey (HRS), a complete volume-limited survey of 323 normal galaxies in the nearby universe \citep{2010PASP..122..261B}. The dust masses for the HRS galaxies were obtained by an empirical recipe based on the SPIRE flux densities, calibrated against modified blackbody SED models and the \citet{2007ApJ...657..810D} dust model \citep{2012A&A...540A..52C}. For both samples, stellar masses are determined from the $i$-band luminosities using the $g-i$ color-dependent stellar mass-to-light ratio relation from \citet{2009MNRAS.400.1181Z}, assuming a \citet{2003PASP..115..763C} initial mass function. The dust-to-stellar mass ratios for our galaxies fall in the range as the HRS galaxies, but tend to populate the upper regions of this plot. In other words, for a given stellar mass, our sample galaxies tend to be rather dustier compared to the general HRS population. 

This is quite surprising in the light of the so-called dust energy balance inconsistency that seems to be applicable for spiral galaxies. For a small number of edge-on spiral galaxies for which detailed radiative transfer fits have been made, the predicted FIR emission from the models generally underestimates the observed FIR fluxes by a factor of about three \citep{2000A&A...362..138P, 2011A&A...527A.109P, 2001A&A...372..775M, 2004A&A...425..109A, 2005A&A...437..447D, 2010A&A...518L..39B, 2012MNRAS.427.2797D, 2012MNRAS.419..895D}. One would therefore expect that dust masses based on SED fits to the observed FIR/submm, such as those for the HRS galaxies in Figure~{\ref{SpecificDustMass.fig}}, would tend to be several times higher than the dust masses based on radiative transfer modeling of optical images, such as the ones we derived for our sample. 

A possible explanation could, again, be the different selection criteria of the two samples. All HRS galaxies are located at a distance between 15 and 25 Mpc, and the late-type galaxy subsample was selected based on a $K$-band flux limit (as a proxy for the stellar mass). Our sample, on the other hand, is more distant and was chosen to show prominent dust lanes, and as such it might be biased to more dust-rich galaxies than the general HRS population. The only way to really test whether this can explain the different location of the samples in Figure~{\ref{SpecificDustMass.fig}} is by calculating the FIR/submm emission that is to be expected from the radiative transfer models for the galaxies in our sample, and comparing it to the observed FIR/submm flux densities (at least for those galaxies for which such data are available). This is beyond the scope of the present paper and will be the subject of a subsequent work.

\section{Conclusions}
\label{Conclusions.sec}

We have selected 12 edge-on, spiral galaxies from the CALIFA survey in order to constrain both their stellar and dust distribution. This was done by computing accurate radiative transfer models to the SDSS $g$-, $r$-, $i$- and $z$-band images simultaneously. As the galaxies are part of the CALIFA survey they already comply to the following criteria: the redshift ranges between $0.005 < z < 0.03$ and the isophotal $r$-band diameter ranges from 45 to 80 arcsec. The galaxies with an obvious dust lane were then selected while avoiding the strongly asymmetrical or interacting ones. As final selection criterium we exclude galaxies with a major axis smaller than 1 arcmin or a minor axis smaller than 8 arcsec as to ensure the dust lane has a high enough resolution to be modeled accurately. 

Before using our automated fitting routine, FitSKIRT, we have tested and validated its capabilities by applying it to a test case described in \cite{2013A&A...550A..74D}. The mock image was created using the radiative transfer code SKIRT, in order to compare both the ability to reproduce the image as well as the recovery of the input values. We found that FitSKIRT was able to give reasonable constraints on all free parameters describing the stellar disc, stellar S\'ersic bulge and dust disc. It is shown that the oligochromatic fitting, i.e. fitting to a number of bands simultaneously, has clear advantages over monochromatic fitting in terms of accuracy. In particular the parameters describing the dust distribution have a smaller spread as the oligochromatic fitting method is less prone to degeneracies in the free parameters. Of the 25000 most central pixels, about 80\% have a value that deviates 20\% or less from the corresponding pixel in the reference image. With these results we can safely apply the method to real data. 

The results of the fits to the 12 galaxies in our sample can be found in Table~\ref{Results.tab} while the sample averages, spread and accuracy can be found in Table~\ref{MeanValues.tab}.

For only two galaxies (UGC\,4163 and NGC\,5908) in our sample, we found that the model, consisting of a exponential disc to describe the stellar and dust distribution and a S\'ersic profile to model the central bulge, was not able to accurately reproduce the observations. In all other cases we were able to model the galaxies and constrain the parameters to an acceptable accuracy. In all of the residual frames more than half of the pixels show deviations of at most 25 \%. Stellar disc and bulge parameters are determined within 10 and 15\% respectively while the dust parameters are less certain, with error bars rising up to 20 or 30\% for the face-on optical depth.  

We find that the average disc scale length and intrinsic disc flattening is in good agreement with the results described by \citet{2002MNRAS.334..646K} and \cite{2009MNRAS.393.1531G}. Our sample, on the other hand, does seem to have larger bulges with an average effective radius of $2.31\pm1.59$~kpc. Possible explanations for this difference are the fact that we do not include a bar in our model and a possible selection effect due to the necessity to have a clear dust lane while \cite{2009MNRAS.393.1531G} does not take into account the effect of dust attenuation on the determined bulge parameters. Consequently a slightly higher bulge-to-total ratio is found although we find a similar trend in the ratios as a function of wavelength. 

For the dust scale length and height we find a good agreement with \citetalias{1999A&A...344..868X} and \citetalias{2007A&A...471..765B}. Also the relative sizes of the dust disc compared to the stellar disc are in good agreement where we find that the dust disc is about 70\% more extended, which is slightly larger than found by \citetalias{1999A&A...344..868X} and \citetalias{2007A&A...471..765B}, but twice as thin as the stellar disc. From \cite{2004ApJ...608..189D} we should expect to see a transition for galaxies with a rotation speed of 120 km\,s$^{-1}$ where the slower rotating ones should have a dust disc scale height similar to the one of the stellar disc. 
Using the baryonic Tully-Fisher relation to get estimates on the rotation speed based on the stellar mass of the galaxies we do not seem to find a similar trend in our sample. A possible explanation might again be that our sample has a clear tendency for larger bulges compared to the bulgeless galaxies investigated by \cite{2004ApJ...608..189D} and \cite{2011ApJ...741....6M}.

An important aspect of the dust distribution in spiral galaxies is the face-on optical depth. A value of higher than 1 means a significant part of the light would be blocked even when the galaxy is seen face-on. In our sample we find a large spread ranging  from 0.18 to 1.98 with an average V-band value of $0.76$ and spread within the sample of $0.60$. As a result, a fraction of the galaxies should not be transparent even when seen completely face-on. This is a slightly larger value in optical depth and in spread than what was previously found by \citetalias{1999A&A...344..868X} and \citetalias{2007A&A...471..765B}. This could be the result of our galaxies requiring a visible dust lane while the galaxies are at larger distance compared to the sample investigated by \citetalias{1999A&A...344..868X} and \citetalias{2007A&A...471..765B}. Therefore the galaxies in our sample are relatively more dust rich than galaxies selected on a dust lane at smaller distances. 

The same difference in distance of this sample and the HRS sample investigated in \cite{2012A&A...540A..52C} could explain why our galaxies reside on the higher side of the relation between the dust-to-stellar mass and the stellar mass. This is an unexpected result, as deriving the dust mass from extinction usually results in values which underestimate the actual dust mass determined from FIR observations by a factor of 3 \citep{2000A&A...362..138P, 2011A&A...527A.109P, 2001A&A...372..775M, 2004A&A...425..109A, 2005A&A...437..447D, 2010A&A...518L..39B}.

A useful way to follow up this research would be to investigate this difference by modeling these galaxies by the means of detailed panchromatic radiative transfer simulations covering the entire SED from the UV to the FIR \citep{2010A&A...518L..39B, 2012MNRAS.427.2797D, 2012MNRAS.419..895D}. Comparing the FIR fluxes predicted from the radiative transfer models determined in this paper with the observed fluxes would yield further insights into the dust energy balance of spiral galaxies.

\section*{Acknowledgements}

G.D.G., M.B., I.D.L.\ and S.V.\ gratefully acknowledge the support of the Flemish Fund for Scientific Research (FWO-Vlaanderen). M.B., J.F.\ and T.H.\ acknowledge financial support from the Belgian Science Policy Office (BELSPO) through the PRODEX project ÒHerschel-PACS Guaranteed Time and Open Time Programs: Science ExploitationÓ (C90370). This work has been realized in the frame of the CHARM framework (Contemporary physical challenges in Heliospheric and AstRophysical Models), a phase VII Interuniversity Attraction Pole (IAP) programme organised by BELSPO, the BELgian federal Science Policy Office. 

Funding for the SDSS and SDSS-II has been provided by the Alfred P. Sloan Foundation, the Participating Institutions, the National Science Foundation, the US Department of Energy, the National Aeronautics and Space Administration, the Japanese Monbukagakusho, the Max Planck Society, and the Higher Education Funding Council for England. The SDSS Web Site is \url{www.sdss.org}. The SDSS is managed by the Astrophysical Research Consortium for the Participating Institutions. The Participating Institutions are the American Museum of Natural History, Astrophysical Institute Potsdam, University of Basel, University of Cambridge, Case Western Reserve University, University of Chicago, Drexel University, Fermilab, the Institute for Advanced Study, the Japan Participation Group, Johns Hopkins University, the Joint Institute for Nuclear Astrophysics, the Kavli Institute for Particle Astrophysics and Cosmology, the Korean Scientist Group, the Chinese Academy of Sciences (LAMOST), Los Alamos National Laboratory, the Max-Planck- Institute for Astronomy (MPIA), the Max-Planck-Institute for Astrophysics (MPA), New Mexico State University, Ohio State University, University of Pittsburgh, University of Portsmouth, Princeton University, the United States Naval Observatory, and the University of Washington. 

This research has made use of the NASA/IPAC Extragalactic Database (NED) which is operated by the Jet Propulsion Laboratory, California Institute of Technology, under contract with the National Aeronautics and Space Administration. This research has made use of NASAÕs Astrophysics Data System bibliographic services. This research has made use of SAOImage DS9, developed by Smithsonian Astrophysical Observatory.

We thank the referee, Dr. E. M. Xilouris for an insightful and timely referee report.

\bibliography{califa}

\end{document}